\documentclass[iop]{emulateapj}

\newcommand{\flux}{erg~cm$^{-2}$s$^{-1}$}
\newcommand{\lx}{erg~s$^{-1}$}

\newcommand{\chandra}{{\it Chandra}}
\newcommand{\xmm}{{\it XMM-Newton}}
\newcommand{\rxte}{{\it RXTE}}

\shorttitle{A Complete Library of X-ray Pulsars in SMC}
\shortauthors{Yang et al.}

\usepackage{amsmath} 
\usepackage{booktabs}
\usepackage [english]{babel}
\usepackage [autostyle, english = american]{csquotes}
\MakeOuterQuote{"}
\usepackage{hyperref}

\newcommand{\RN}[1]{%
  \textup{\uppercase\expandafter{\romannumeral#1}}}

\begin{document}

\title{A Comprehensive Library of X-ray Pulsars in the Small Magellanic Cloud: \\ Time Evolution of their Luminosities and Spin Periods}

\author{J. Yang\altaffilmark{1,2,3}, S. G. T. Laycock\altaffilmark{1,2}, D. M. Christodoulou\altaffilmark{1,4}, S. Fingerman\altaffilmark{1,2}, M. J. Coe\altaffilmark{5} and J. J. Drake\altaffilmark{3}} 

\altaffiltext{1}{Lowell Center for Space Science and Technology, University of Massachusetts, Lowell, MA 01854. }
\altaffiltext{2}{Department of Physics and Applied Physics, University of Massachusetts,
    Lowell, MA 01854. }
\altaffiltext{3}{Harvard-Smithsonian Center for Astrophysics, Cambridge, MA 02138. Email: jun.yang@cfa.harvard.edu}
\altaffiltext{4}{Department of Mathematical Sciences, University of\\ Massachusetts, Lowell, MA 01854. }
\altaffiltext{5}{Physics \& Astronomy, University of Southampton,
    SO17 1BJ, UK. }

\begin{abstract}
We have collected and analyzed the complete archive of  {\itshape XMM-Newton\/} (116), {\itshape Chandra\/} (151), and  {\itshape RXTE\/} (952) observations of the Small Magellanic Cloud (SMC), spanning 1997-2014. The resulting observational library provides a comprehensive view of the physical, temporal and statistical properties of the SMC pulsar population across the luminosity range of $L_X= 10^{31.2}$--$10^{38}$~erg~s$^{-1}$.  From a sample of 67 pulsars we report $\sim$1654 individual pulsar detections, yielding $\sim$1260 pulse period measurements. Our pipeline generates a suite of products for each pulsar detection: spin period, flux, event list, high time-resolution light-curve, pulse-profile, periodogram, and spectrum. Combining all three satellites, we generated complete histories of the spin periods, pulse amplitudes, pulsed fractions and X-ray luminosities. Some pulsars show variations in pulse period due to the combination of orbital motion and accretion torques. Long-term spin-up/down trends are seen in 12/11 pulsars respectively, pointing to sustained transfer of mass and angular momentum to the neutron star on decadal timescales. Of the sample 30 pulsars have relatively very small spin period derivative and may be close to equilibrium spin. The distributions of pulse-detection and flux as functions of spin-period provide interesting findings: mapping boundaries of accretion-driven X-ray luminosity, and showing that fast pulsars ($P<$10 s) are rarely detected, which yet are more prone to giant outbursts. Accompanying this paper is an initial public release of the library so that it can be used by other researchers. We intend the library to be useful in driving improved models of neutron star magnetospheres and accretion physics.
\end{abstract}

\accepted{16 February 2017}


\section{Introduction}\label{intro}

The purpose of this paper is to motivate and facilitate the community to investigate on a population-statistical basis, the processes of accretion and X-ray emission in High Mass X-ray Binary (HMXB) pulsars. To this end we have applied a uniform reduction/analysis pipeline to compile a comprehensive set of physical  parameters from the large archive of observations of the Small Magellanic Cloud (SMC). 

The SMC is a dwarf irregular galaxy at a distance of 62 kpc from the Milky Way \citep{gra14, sco16}. It contains a large and active population of HMXBs \citep[e.g.][]{gal08,tow11,klu14,coe15,hab16,chr16}. Systems involving a Be type star account for 98\% of the confirmed SMC HMXBs \citep{coe05}. 

Be/X-ray binaries (Be-XBs) are stellar systems in which a neutron star (NS) accretes matter from the circumstellar disk of a massive early-type companion resulting in outbursts of high-energy radiation. During all such outbursts the X-ray flux is pulsed at the spin period of the neutron star, due to magnetic channeling of the accretion flow to the poles of the neutron star, hence these objects are transient X-ray pulsars. The triggering of outbursts is partly controlled by the orbital parameters of the binary, with many systems exhibiting strings of sub-Eddington outbursts spaced at the orbital period, occurring close to periastron, and lasting for about 1/10 of the orbit. Not all periastron passages lead to an outburst so other factors must be involved, and are the subject of current research. Some Be-XBs have undergone infrequent but much more luminous outbursts, exceeding the Eddington luminosity ($L_{Edd}$) and lasting for a complete orbit or longer (e.g., \citealt{Martin2014}). The cause of the latter \textquotedblleft{giant outbursts}" is suspected to be due to a sudden increase in the mass-transfer rate from the companion, driven by tidally induced density perturbations in the circumstellar disk \citep{Neg1998,Moritani2013}. These evolving dynamic structures are observed in the H$\alpha$ line profiles of Be-XBs \citep{Reig2016}. It is also possible that a radiative instability in the pulsar's accretion disk, similar to the situation leading to a nova eruption, plays a key role in triggering giant outbursts \citep{Okazaki2013}. 

Proximity, compact size, and minimal foreground contamination have made the SMC an ideal laboratory to study HMXBs. The motivation of this paper is to advance the study of its population as a whole, to enable robust statistical analysis of the physical parameters of HMXBs, and in turn enable a new generation of theoretical models to improve understanding of accretion physics.  The X-ray source population of the SMC is different than that found in the Milky Way, the Large Magellanic Cloud (LMC), and other Local Group galaxies (e.g., M31, M33). Many more HMXBs are known in the SMC than in the LMC and the Milky Way \citep{hab00, yok03}. Based on the masses of the Milky Way and the SMC, the SMC HMXBs are a factor of 50 more numerous than what one would expect \citep{wal06, mcb08}. The SMC is experiencing an era of ongoing star formation and its large number of HMXBs is certainly related to the high star formation rate (SFR) \citep{gri03}. The timescale for production of HMXBs in the SMC has been revealed by \cite{Antoniou2010} to peak in the 25-60 Myr range, making the pulsars in our sample of similar age. The above factors establish the SMC as a unique place to study this important branch of stellar evolution. 

Building on the legacy of almost 2 decades of dedicated monitoring of the SMC, in this paper we study observations collected in surveys carried out with {\itshape XMM-Newton\/} \citep{hab08}, {\itshape Chandra\/} \citep{ant09, lay10}, and {\itshape RXTE\/} \citep{lay05, gal08}. The basis of our analysis is the SMC X-ray pulsar (SXP) catalog of \cite{coe15}. This catalog lists the coordinates, orbital period, eccentricity, measured spin period of the compact object, and characteristics of the companion Be star where known for currently identified SMC X-ray pulsars. We have mined the data from the archives of these three telescopes and we have generated a new comprehensive time-domain library of high-level data products for the SMC pulsars in our resulting sample.  

This new archive incorporates all previous surveys of the SMC and adds all the latest observations that fall in the public domain up to the year 2014. 
For each known SMC pulsar, all of these archived data are combined together in order to produce a complete picture of the X-ray emission.
The library also contains comprehensive (folded and unfolded) light curves at different energy bands, the variations of the luminosities, the pulsation amplitudes (count rate), the pulsed fractions, and $\dot{P}$ information for each known pulsar, all of which portray the long-term behaviors of these objects. 

The resulting products can be used toward making progress in the following areas:
(a)~map out the duty cycles of X-ray emission from HMXBs and delineate the various phases of accretion and quiescence;
(b)~provide a library of pulse profiles to confront geometric models \citep[e.g.][]{yan17} of pulsar emission and constrain their physical and geometrical parameters; 
(c)~investigate statistical correlations between the physical  properties of the compact objects, across the full parameter space;
and
(d)~produce large-number statistics for the entire class of objects.

This wealth of information will also be released for public use. Release 1.0 (coincident with the publication of this paper) comprises the catalog of measurements obtained from our pipeline processing. The scope and contents of the available database is illustrated by the plots and tables presented throughout this paper.  Release 2.0 will include pulse profiles and photon event files. 

In the following sections we describe the contents of the new library, how 
the information is grouped, and how it can be used.
In \S~\ref{obs}, we describe the observations from the three observatories and our processing of the raw data from each archive. In \S~\ref{sect:library}, we describe the products and content of the library, illustrated with examples of individual  pulsars, and present some statistical inferences from the combined data set of the 3 satellites and SMC pulsars. We conclude in \S~\ref{summary} with a discussion of our products and a summary of the paper.

\section{Observations}\label{obs}

The library is constructed from a large volume of archival {\it XMM-Newton, Chandra} and {\it RXTE} data, which is summarized in Table~\ref{tab:observations} and described in detail in the following subsections.    

\begin{deluxetable*}{l|lll}
\tablecaption{Summary of All Pulsar Observations} 
\tablehead{ \colhead{Attribute} & \colhead{RXTE} & \colhead{Chandra-ACIS }    & \colhead{XMM-Newton-PN}    \\ }
\startdata
No. ObsID.s                              &   952        &    151    &    116        \\
Date Range 				& 1997/11/25 - 2011/12/24  & 1999/08/27 - 2014/01/15 & 2000/04/16 - 2014/10/20 \\
Date Range  (MJD)                             &  50777-55919       &   51417- 56672    &     51650-56950       \\
No. of Obs.$^1$  		& 36316     & 517  &  326       \\
No. Detections$^2$      &    n/a       & 232    &  254     \\
No. Pulsations$^3$     &  1113      & 69    &   78    \\
\enddata
\tablecomments{Summary of the contents of the Pulsar Library by Satellite.
(1)Sum of: number of pulsars in FOV per ObsID. 
(2) For imaging instruments only, the number of detections of point-sources at known pulsar coordinates. 
(3) The total number of pulsar observations in which a fundamental spin period was detected at $\geqslant99\%$ significance.}
\label{tab:observations}
\end{deluxetable*}

\begin{figure*}[h]
\epsscale{1.00}
\plotone{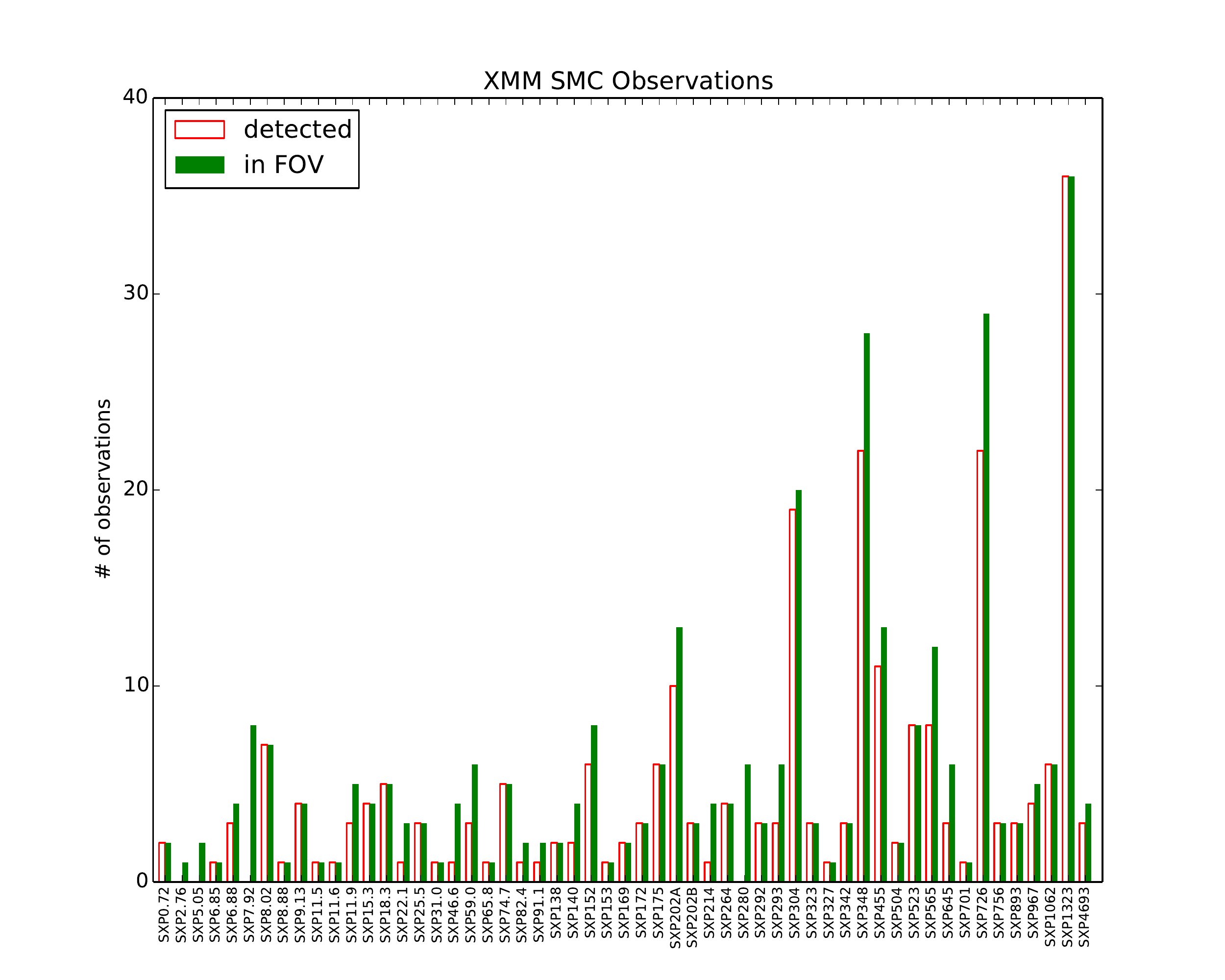}
\caption{Number of {\itshape XMM-Newton\/} observations of each known pulsar in the SMC. The pulsars are arranged in order of increasing spin period in seconds, as their names indicate. Filled green bars show how many times each pulsar was in the FOV; unfilled red bars show how many times each pulsar was actually detected by {\itshape XMM-Newton\/}. }
\label{fig:xmm}
\end{figure*}

\subsection{\textit{XMM-Newton} Observations}

\textit{XMM-Newton} was launched in December 1999 by the European Space Agency. It carries three identical grazing incidence X-ray telescopes, whose point spread function (PSF) is sufficient to spatially separate most of the individual bright X-ray sources of the SMC. 
Up until 2014 (the cutoff for this project), 116 \textit{XMM-Newton} observations of the SMC were available in the archive, as listed in Table~\ref{tab:observations}.

\textit{XMM-Newton} 
has the largest effective area of any X-ray satellite at energies below 2~keV, a record it will retain until the Neutron Star Interior Composition ExploreR (NICER
) launches; it also has a 30~cm optical/UV telescope \citep[the Optical Monitor;][]{mas01}
allowing simultaneous X-ray and optical/UV coverage. \textit{XMM-Newton}'s X-ray telescopes each feed one of the European Photon Imaging Cameras (EPIC) \citep{str01,tur01}, which comprise the PN, and Metal Oxide Semi-conductor (MOS1, MOS2) instruments covering the energy range 0.1--15~keV with moderate spectral resolution. 

From the {\itshape XMM-Newton\/} archive we acquired the EPIC PN data (which have a higher time resolution than the MOS data) for all publicly available SMC observations obtained from 2000 to 2014. 
In Fig.~\ref{fig:xmm} the number of observations in which the different known SMC pulsars appeared in the field of view (FOV) and the number of positive detections of them by the PN camera are shown in green and red, respectively. The spin periods of these SMC pulsars range from 0.72 to 4693 s. Here positive detections means photon counts and flux are recorded in the \textit{XMM-Newton} Science Archive (XSA)\footnote{\url{http://nxsa.esac.esa.int/nxsa-web/}}. The X-ray source detections are above the processing likelihood of 6\footnote{\url{http://xmmssc.irap.omp.eu/Catalogue/3XMM-DR6/3XMM-DR6_Catalogue_User_Guide.html}}.
In this histogram, we can see how many times each source was caught in quiescence (corresponding to non-detections)
and we can identify the sources that are appear to be permanent (at least in the context of our 15 year study baseline) rather than transient emitters.

In our data analysis pipeline, we begin with the 
SXP catalog of \cite{coe15} which contains pulse periods and celestial coordinates of each known pulsar, determined from the existing body of publications which includes X-ray and optical counterpart identifications. Most of the SXP pulsars have positions known to sub-arcsecond accuracy.  We take the {\itshape XMM-Newton\/} source catalog \cite{} obtained from the XSA and search for EPIC detections (both MOS and PN) in a positional search radius for each known pulsar. The initial search uses a search radius of 15$''~$about the known SXP coordinates. Every XSA catalog point-source within this radius is evaluated based on its positional offset and uncertainty, requiring the offset to be smaller than 3$\sigma_c$ for a positive identification with the SXP object, where $\sigma_c$ is the combined uncertainty computed via Equation ~\ref{eqn:errorcirc},
\begin{equation}
\label{eqn:errorcirc}
\sigma_c \equiv \frac{r_{off}}{\sqrt{r_{xmm}^2+ r_{sys}^2 + r_{psr}^2}},
\end{equation} 
and where $r_{off}$ is the offset between the known position and the detected point source position, $r_{xmm}$ is the {\itshape XMM-Newton\/} Science Archive (XSA) Right Ascension and Declination (RADEC) combined error (determined while fitting the detection), $r_{sys}$ is the XSA systematic error of the {\itshape XMM-Newton\/} fields, and $r_{psr}$ is the uncertainty of the known position of the pulsar.  

Our data reduction of these sources included the standard procedures of the {\itshape XMM-Newton\/} Science Analysis Software (SAS, version 1.2) from the {\itshape XMM-Newton\/} Science Operations Center (SOC). We retained the standard {\itshape XMM-Newton\/} event grades (patterns 0-4 for the PN camera), as these have the best energy calibration
. The PN data were analyzed using the commands \textit{evselect} and \textit{epiclccorr} in SAS. 
For the EPIC detectors, (Str\"{u}der et al. 2001; Turner et al. 2001), 
the data for the point-like sources were extracted from circular regions of radius 20$''$.  Background levels were estimated and subtracted using annular regions defined by inner and outer radii of 30$''$ and 60$''$ centered on each source.

Source fluxes were directly obtained from the XSA using the known pulsar coordinates. If a source was not detected but it was in the FOV, then we recorded the upper limit to the flux that was calculated from the \textquotedblleft{Flux Limits from Images from XMM-Newton}" (FLIX)\footnote{\url{http://www.ledas.ac.uk/flix/flix3}} server.
 
It was necessary to extract \textit{XMM-Newton} light curves from scratch at the native timing resolution of the detectors (0.0734~s for PN) because the XSA pipeline versions are generated using a binning scheme that precludes high-resolution timing analysis. We also extracted and archived the event list (Time, Energy) for each individual source. This product is the starting point for performing a range of advanced event-based analyses, e.g. FFT~\citep{IS1996}, quantiles~\citep{hon05}, energy dependent pulse profiles and 2D phase-energy-intensity histograms \citep{Sch2014,Hong2016}.   

The times of arrival of the photon events were shifted from the local satellite frame to the barycenter of the solar system using the task {\itshape barycen} in the SAS software. 
Additional improvements using the cleaned event file and the \textquoteleft{gti}' (good time interval) task \textit{tabgtigen} were also made in order to exclude the times when the background was very high. The rate threshold of the \textquoteleft{gti}' \textit{fits} file applied to the event file was set at 0.6 counts/s.
A pulsation search was performed following the procedure described in Section~\ref{sect:pulsesearch} to search for periodicities in the broad, soft (0.2--2 keV), and hard (2--12 keV) energy bands. 


\begin{figure*}[h]
\epsscale{1.00}
\plotone{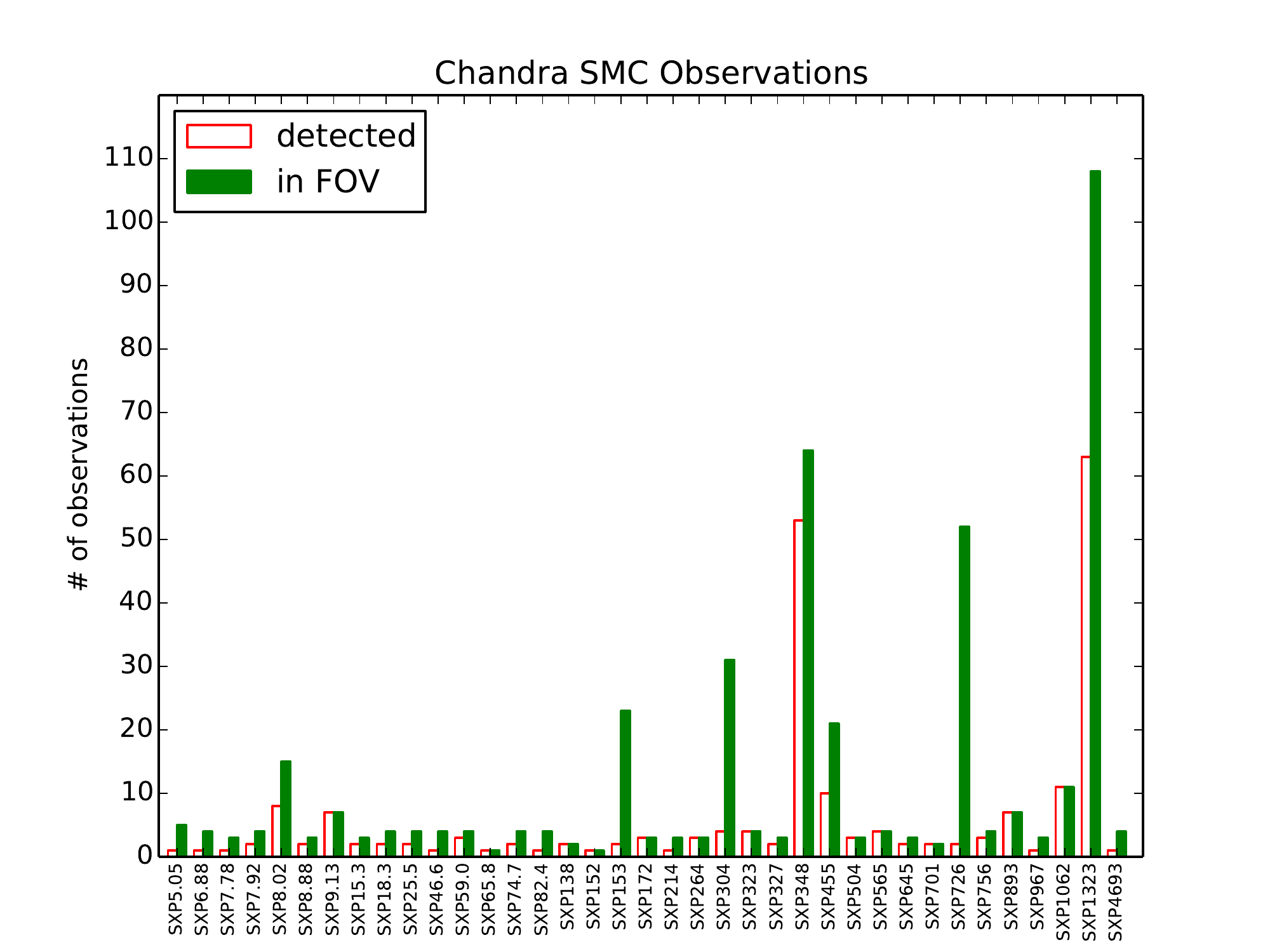}
\caption{Number of {\itshape Chandra\/} observations of each known pulsar in the SMC. The pulsars are arranged in order of increasing spin period in seconds, as their names indicate. Filled green bars show how many times each pulsar was in the FOV; unfilled red bars show how many times each pulsar was actually detected by {\itshape Chandra\/}.
}
\label{cha}
\end{figure*}

\subsection {\textit{Chandra} Observations}

The Chanda X-ray Observatory was launched in July~1999, carrying the High Resolution Mirror Assembly (HRMA), the highest angular-resolution X-ray optics ever flown in space
. The HRMA focusses X-ray light on either the Advanced CCD Imaging Spectrometer (ACIS
) or the High Resolution Camera (HRC
). For this project we used ACIS data, which provide sub-arcsecond source positions, photon energies, and event timing to 3.2~s resolution in standard operation. 

The known X-ray pulsars in the SMC \citep{coe15} were searched for in the first fifteen years of \textit{Chandra} observations obtained from the \textit{Chandra} Data Archive.  We applied the latest \textit{Chandra} calibration files to each image and we reduced the data with the {\itshape Chandra\/} Interactive Analysis of Observations software package (CIAO, version 4.5) \citep{fru06}.  We created X-ray images within the 0.3--7.0 keV energy band in which \textit{Chandra} is best calibrated and most sensitive. First, the image files were obtained by executing \textit{fluximage} on the reprocessed event files. We then used \textit{mkpsfmap} to generate a corresponding image whose amplitude is the PSF size in terms of the region that encloses 95\% of the counts at 1.5~keV. The task \textit{evalpos} was used to look up the PSF size at each source position in this image, and \textit{dmmakereg} was used for generating the appropriate sized circular source extraction and annular background extraction regions.

After verifying the accuracy of the \textit{Chandra} astrometry for the fields of interest, we placed our extraction regions at the coordinates of each pulsar to extract the source and background events.  
Finally, source fluxes and light curves were extracted with the CIAO tools \textit{srcflux} and \textit{dmextract}, respectively. The tool \textit{srcflux} was used with a power-law spectral model with photon index 1.5 and an absorption column $5 \times 10^{21}$~cm$^{-2}$, representative of the line-of-sight to the SMC. 

\begin{figure*}[h]
\epsscale{1.00}
\plotone{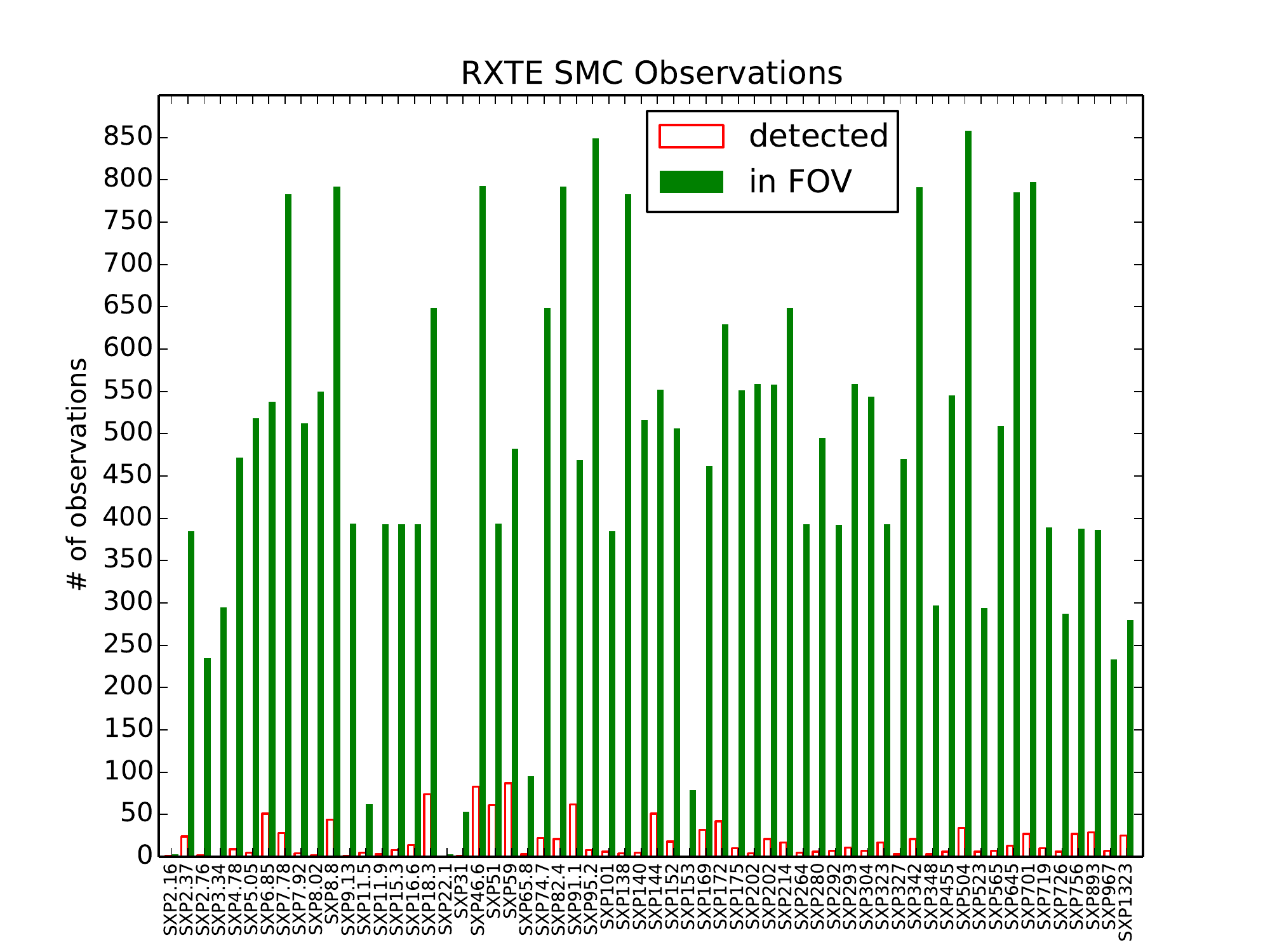}
\caption{
Number of {\itshape RXTE\/} observations of each known pulsar in the SMC. The pulsars are arranged in order of increasing spin period in seconds, as their names indicate. Green bars show how many times each pulsar was in the FOV with the observations of collimator response $>$ 0.2; unfilled red bars show how many times each pulsar with $\geqslant$ 99\% significant pulsations was actually detected by {\itshape RXTE\/} with collimator response $>$ 0.2.
}
\label{rxt}
\end{figure*}

\textit{srcflux} was also used to determine whether the pulsar was detected, in which case a flux measurement is reported, or not, in which case an upper limit is reported instead. The source flux was extracted within the region that encloses 95\% of the counts at 1.5 keV. This region was determined for each detection seperately using the PSF map. The {\it srcflux} derived net photon counts $\geqslant 1$ at 90\% (1.65 $\sigma$) confidence level were recorded as detections. If the net counts are 0, then the upper limit is reported. For {\it srcflux} positive detections, a source event file, light curve, and pulse-height spectrum were extracted, and saved in the library along with the appropriate response files. Light curves were extracted at 3.2 s resolution, set by the read-mode most commonly encountered in the archival dataset. A pulsation search was performed following the procedure described in Section~\ref{sect:pulsesearch}.

The number of times each source was in a \textit{Chandra} FOV and the number of times each source was detected by {\itshape Chandra\/} are shown in Fig.~\ref{cha}. As in the analogous Fig.~\ref{fig:xmm} for the \textit{XMM-Newton} data, this plot serves to illustrate the frequency of each source being caught in quiescence (non-detections) and allows us to identify the ``permanent'' rather than transient emitters.

\subsection{{\itshape RXTE\/} Observations}

The Rossi X-ray Timing Explorer (\textit{RXTE}
) launched in December 1995 carrying the Proportional Counter Array (PCA), All-Sky Monitor (ASM), and High Energy X-ray  Timing Experiment. The PCA consisted of 4 Xenon-filled Proportional Counter Units (PCUs), and provided individual event timing at $\mu s$ resolution and moderate energy resolution over the range 2--60 keV.  Starting in 1997 the SMC was observed approximately weekly with the PCA for some 16 years, accumulating $\sim$1000 observations (Table~\ref{tab:observations}) before \textit{RXTE} was deactivated in 2012 January.  
Data in Good Xenon mode were extracted in the 3--10 keV energy range (maximizing S/N) using the FTOOLS suite. 
In our subsequent analysis, we used the output of the IDL pipeline products (PUlsar Monitoring Algorithm or PUMA) up to year 2012 \citep{gal08}. To summarize the data reduction: the data were first cleaned using standard FTOOLS scripts\footnote{\url{http://heasarc.gsfc.nasa.gov/ftools/}}, then the FTOOL \textit{maketime} was used to generate the GTI files. We extract the light curves with 0.01~s binning\footnote{\url{https://heasarc.gsfc.nasa.gov/docs/xte/abc/extracting.html}} and we applied background subtraction and barycentric correction. Finally, the count rates were normalized to account for the varying number of active Proportional Counter Units. Further details can be found in \citealp{gal06, gal08, tow11, tow13, klu14}. 

The number of \textit{RXTE} observations with collimator response $>$ 0.2 of each known SMC pulsar is shown in Fig.~\ref{rxt}. \textit{RXTE} PCA was not an imaging detector, and this figure shows only the number of observations for which the coordinates of each pulsar were in the 2$^{\circ}$ Full Width at Zero Intensity (FWZI) FOV. 

\subsection{Data Processing}
\subsubsection{Pulsar Detection}
\label{sect:pulsesearch}

Following the approach established in prior works based on the Lomb-Scargle (LS) periodogram  (\citealt{lay05, lay10, gal08}), we searched for pulsations in the light curves of each pulsar from each satellite detection. 
The significance, $s$, of each periodicity was calculated from the number {\itshape M\/} of independent frequencies and the LS power $P_X$  according to \citet{pre92}, 
\begin{equation}
s=[1-M \exp(-P_X)]\cdot 100\% .
\end{equation}
The error in spin period was calculated from the standard deviation of the frequency \citep{hor86},
\begin{equation}
\delta \omega= \frac{3\pi \sigma_N}{2 T A_s \sqrt{N}} ,
\end{equation}
where
$\sigma_N ^{2}$ is the variance of the light curve, {\itshape N\/} is the number of data points, {\itshape T\/} is the total length of the data, and $A_s$ is the amplitude of the signal. 

Each periodogram was automatically scanned to look for the expected fundamental and harmonics of the pulsar. For the imaging instruments the specific pulsar is known, and so the search is very targeted. For {\it RXTE} the periods of all known pulsars in the FOV of the PCA collimator are searched, and we used the fundamental harmonics of $[0.8-1.2] \times P$ with the $s\geqslant99\%$ and all the collimator responses in the later analysis. Confidence levels are assigned following the prescription of \cite{pre92} computed for a search over a 10\% tolerance range on the expected period. We treat significance $\geqslant$99\% as a valid detection and do not adjust the threshold to account for the number of observations. This ensures a uniform criterion that does not change if the definition of the sample were to change. 
When a detection was made we obtained the spin period, then folded the light curve and obtained the pulsation amplitude and the pulsed fraction (PF). The pulsed fractions of the light curves were calculated by integrating over the pulse profile according to the prescription of \cite{bil97}. Here

\begin{equation}
PF=\frac{\frac{\sum_{j}^{ n_{bin}} (f_{j,mean}-f_{min})}{n_{bin}}}{\frac{\sum_{i}^{N } f_{i}}{N}}, 
\end{equation}
where $n_{bin}$ is the number of bins for each folded light curve, $f_{j,mean}$ is the mean photon count rate in each bin, $f_{min}$ is the minimum of $f_{j,mean}$, and $f_{i}$ is the photon count rate of the un-binned light curves.

Note that for {\itshape RXTE\/} we did not compute pulsed fractions since the {\it PCA} is a non-imaging detector and multiple sources are always in the FOV, so the un-pulsed component cannot be reliably measured.

\subsubsection{Cross Calibration}
\label{sect:calibration}
In order to combine the data for the three satellites together it is necessary to account for differences in sensitivity and energy range. In principle the response functions of each instrument are well characterized and good cross calibration for flux and luminosity is possible for the imaging instruments. Since RXTE is not an imaging instrument, and has very different properties, we took an empirical approach to scale the pulse amplitudes. For the purposes of plotting our results in this paper we performed a cross-calibration to normalize the luminosities and pulse amplitudes (count rates) to be on the same scale as the {\itshape XMM-Newton\/} PN. In the case of Chandra this was accomplished using the tool PIMMS\footnote{\url{http://cxc.harvard.edu/toolkit/pimms.jsp}} (Portable, Interactive, Multi-Mission Simulator) to
convert the count rate from the ACIS-I detector (with no grating) into {\itshape XMM-Newton\/} PN count rate (with no grating and medium filter). The input energy range was 0.3--7 keV and the output energy range was 0.2--12 keV. 
We used power-law models with absorption $5 \times 10^{21}$~cm$^{-2}$ (calculated using NASA's HEASARC tool\footnote{\url{https://heasarc.gsfc.nasa.gov/cgi-bin/Tools/w3nh/w3nh.pl}} for the SMC sightline) and photon indices of 1.5. With a 1 count/second {\itshape Chandra\/} detection, the predicted {\itshape XMM-Newton\/} count rate is about 4 counts/s. Thus, the pulsation amplitudes from {\itshape Chandra\/} were scaled by a factor of 4. 

 
For the {\itshape RXTE\/} data we do not directly measure absolute flux due to the likelihood of source confusion. Many observations feature 2 or more pulsars active simultaneously (commonly 1--3, but in one case 7).  Instead we followed an empirical approach to estimate total flux from pulse amplitude (which we measure directly): we used PIMMS to calculate the absorbed fluxes from the measured pulse amplitudes, assuming a fixed pulsed fraction. We set both the input and the output energy range of the PCA detector to 3-10 keV. From the PIMMS power-law model described above, the predicted flux for 1 count/PCU/s was ${\cal F}=(9.23_{+0.10}^{-0.12})\times 10^{-12}~\text{erg}~ \text{cm}^{-2}~\text{s}^{-1}$. A proxy for the X-ray luminosity was then calculated according to  
\begin{equation} \label{Lumi-rxte}
\begin{split}
L_X &  = \frac{A~{\cal F}}{PF}~4 \pi  r^2 \\
      & = ({1.06\times 10^{37}}  ~ \text{erg~s}^{-1}) A ,
\end{split}
\end{equation} 
where $A$ is the pulsation amplitude in the unit of counts/s, $PF$ is the (unknown) pulsed fraction which can vary between 0.1 and 0.5 \citep{2015MNRAS.447.2387C}, and $r=62$~kpc is the distance to the SMC.
In the last step of eq.~(\ref{Lumi-rxte}), we chose to set $PF=0.4$ so that our X-ray luminosities are
comparable with the corresponding values obtained by \cite{klu14} who used average count rates (over many observations) instead of individual $A$ values to calculate fluxes. 
\begin{figure*}
\epsscale{1.00}
\plotone{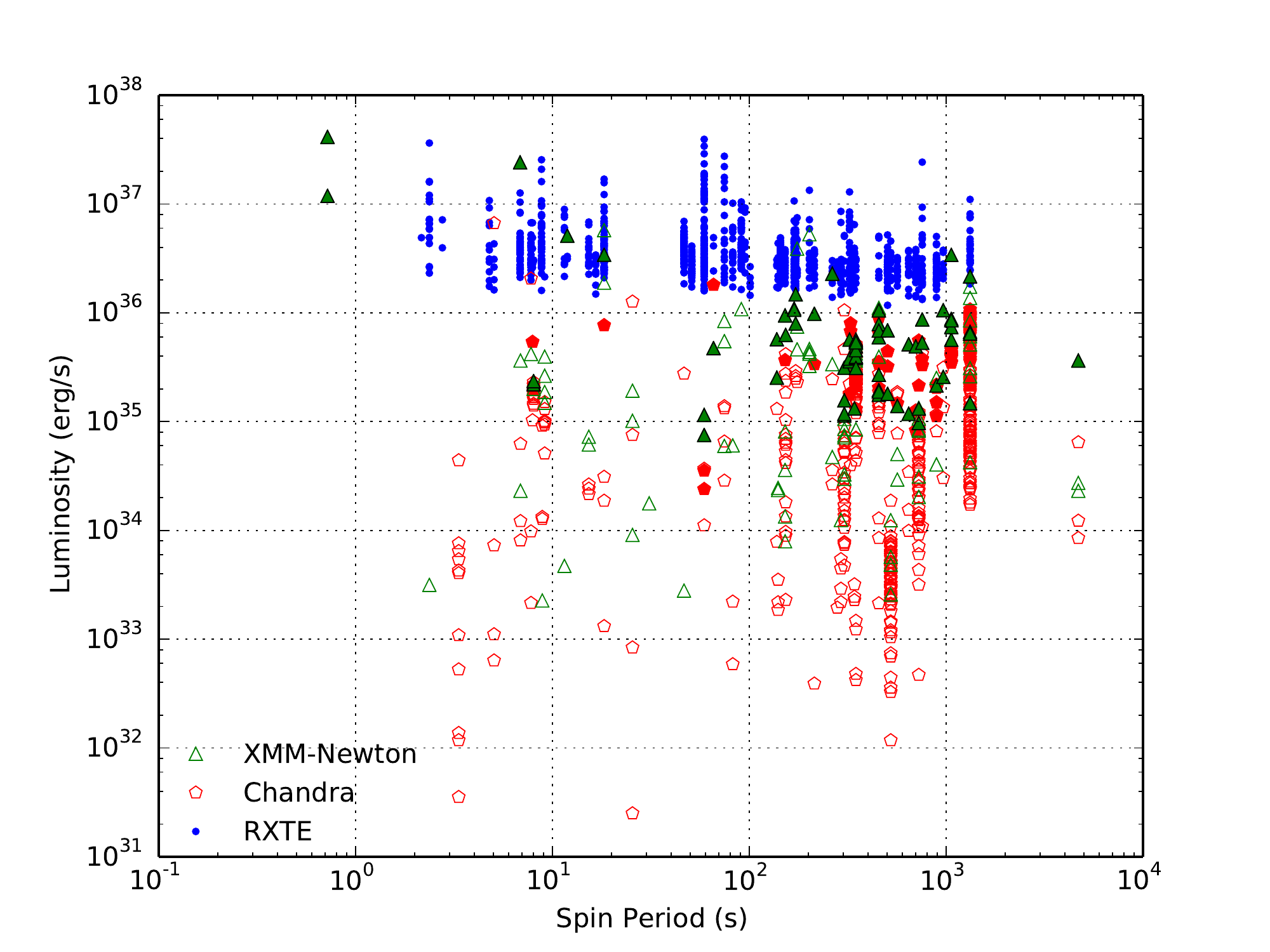}
\caption{$L_X$ vs. $P$ of X-ray pulsars in the SMC from {\itshape XMM-Newton\/}, \textit{Chandra}, and \textit{RXTE} observations. Blue symbols are the \textquoteleft{proxy}' luminosities derived from significant ($\geqslant$99\%) pulse amplitude detections in \textit{RXTE} data (Equation~\ref{Lumi-rxte}).  Filled (Unfilled)
symbols indicate that pulsations are (not) detected in each particular
observation.
} 
\label{fig:lumi}
\end{figure*}

\section {Overall Properties of the SMC Library}\label{sect:library}

Our library includes data products from the \textit{XMM-Newton}, \textit{Chandra} and \textit{RXTE} satellites processed using our data reduction and analysis pipelines, together with parameters computed from the combined products. The number of times that each pulsar was observed by each satellite is reported in Table~\ref{tab:pulsars}, together with the number of such observations that yielded a positive detection (see also Figs. 1-3), so that the reader can readily grasp the extent of the library.  There are several possible outcomes of an observation at the location of a known pulsar: (1) no detection, (2) detection of a point source (for \textit{XMM-Newton} and \textit{Chandra} only), (3) detection of pulsation corresponding to the expected pulse period and/or its harmonics. We have enumerated the number of occurrences of each type of outcome in Table~\ref{tab:observations} for the survey as a whole, and by satellite. 
There are in total $\sim$37,000 observations of pulsars, yielding 1599 individual detections in the library, of which 1260 show pulsation and can be expected to yield pulse profiles. The detailed breakdown by individual pulsar is provided in Table~\ref{tab:pulsars}.  

For each known SMC pulsar detection in the  \textit{XMM-Newton} and \chandra~observations, we have extracted single source event lists\footnote{the XSA source catalog products do not include source event files}, 
spectra, source (background and background-subtracted) light-curves, periodograms, and folded light-curves in the broad (0.2--12), soft (0.2--2 keV), and hard (2--12 keV) energy bands.  Similar products have been produced for \textit{RXTE} with a few differences (single energy-band 3--10 keV light curves which are generated on a per-observation rather than per-pulsar basis, no individual source event files). We have combined the three satellite detections together, to explore the long term behavior of each pulsar. Of course, some sources have been detected by one, two or all three satellites. 

In the following sections we present an overview of the multi-satellite results: (Section~\ref{sect:Lx}) presents the distribution of count-rates and $L_X$; Section~\ref{sect:history} provides examples of the history and time evolution of properties for individual sources and long-term period derivatives; Section~\ref{sect:Pdot} shows how pulse period variations can be analyzed for each individual pulsar; and Section~\ref{sect:access} describes the online catalog and public access to the library. 



\subsection{Luminosity}
\label{sect:Lx}
Figure~\ref{fig:lumi} shows the X-ray luminosities from all observations we mentioned and from all three telescopes. 
The observations cover the range $L_X= 10^{31.2}$--$10^{38}$~erg~s$^{-1}$. 
Filled and unfilled symbols indicate that pulsations were detected or not detected, respectively.
We see that \textit{RXTE} has provided the majority of the detections, since it has observed the most frequently (approximately every week for 15 years). The additional value of \chandra~and \xmm~is clear in two ways. Firstly these imaging instruments are able to distinguish pulsed emission from un-pulsed. Secondly they are in the form of many detections at lower luminosity, which increases the dynamic range. This is an important attribute because the transition between different accretion modes likely occurs below the typical sensitivity of \textit{RXTE}.  
Figure~\ref{fig:lumi} provides valuable information on the disappearance of pulsation. And in this context 
we have recently performed an analysis of the propellor transition in this ensemble of pulsars \citep{chr16}. Minor differences between \citealt{chr16}  and our Fig.~\ref{fig:lumi} are due to inclusion of XMM-MOS data in the former work, and slightly different S/N screening criteria.


In terms of detecting pulsations, a useful empirical result on the sensitivities of \xmm, \chandra~and \rxte~to pulsars as a function of count rate was obtained by analyzing the library. By constructing histograms of source counts for all positive pulsation detections ($s\geqslant99\%$) the distributions of pulsation amplitude, and number of events per detection can both be examined. 

\begin{figure}
\includegraphics[scale=0.45]{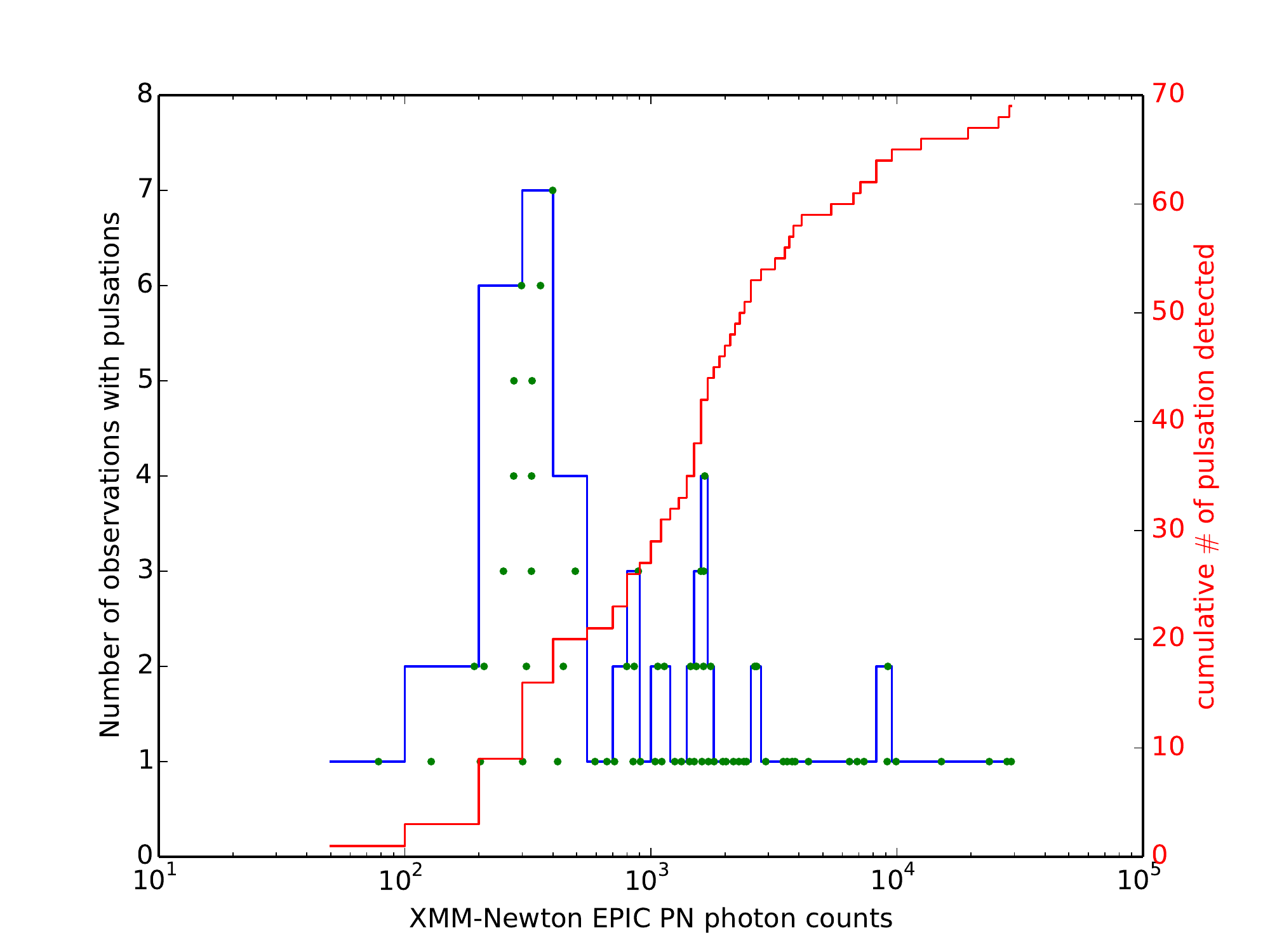}
\caption{Distribution of number of times pulsations are detected ($s \geqslant$ 99\%) as a function of source counts observed by {\itshape XMM-Newton\/}. Green dots are the detected photon counts for each pulsar.}
\label{count}
\end{figure}

Figure~\ref{count} shows the number of observations with pulsations detected by {\itshape XMM-Newton\/} at $s\geqslant 99\%$. The abscissa shows the actual EPIC PN photon counts that yielded the pulsation detections in the histogram. The ordinate of the blue histogram is the  number of sources from the observations with pulsations detected in each bin (in intervals of 100 counts). The distribution indicates a threshold of about 200 counts below which our ability to detect pulsations becomes rapidly diminished. The distribution of detected pulsations peaks at around 300 counts for \textit{XMM-Newton} and represents the completeness limit of the survey. Observations should therefore be designed to obtain 200+ counts when searching for periodicities. The decline towards lower net counts is due to the expected decline in ability to resolve pulsations in fainter sources and shorter observations, while the decline toward higher net counts reflects the underlying distribution of pulsar luminosity. The red histogram is the cumulative number of observations with pulsations detected as a function of photon counts. In total, there are $\sim$70 {\itshape XMM-Newton\/} observations with pulsations detected in the SMC.

\begin{figure}
\includegraphics[scale=0.45]{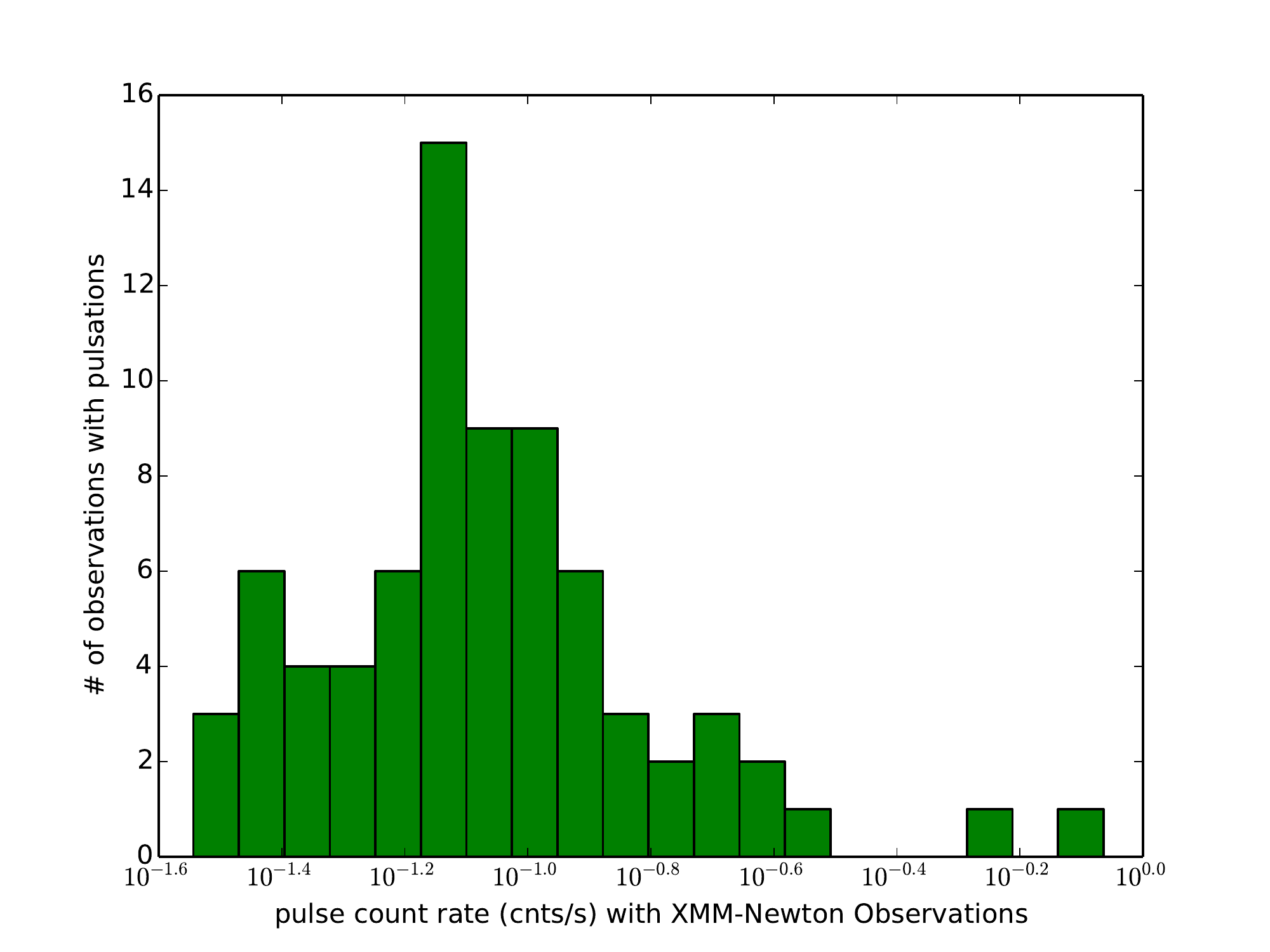}
\caption{Histogram of the number of {\itshape XMM-Newton\/} observations in which pulsations were detected with $s \geqslant$ 99\% as a function of pulse amplitude in terms of count rate binned on equal logarithmic intervals.}
\label{hisxmm}
\end{figure}

\begin{figure}
\includegraphics[scale=0.45]{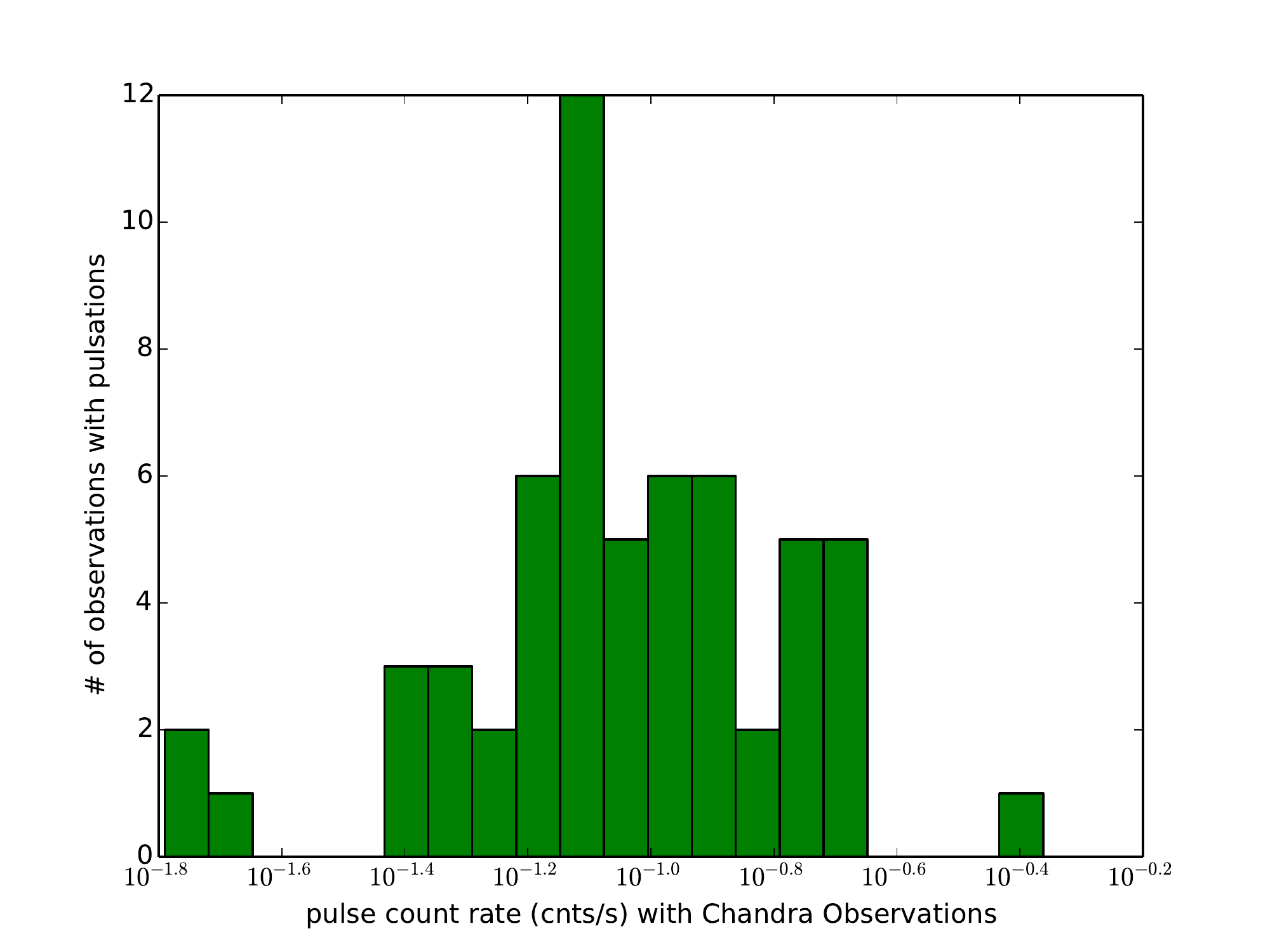}
\caption{Similar to Fig.~\ref{hisxmm}, showing the frequency histogram of pulse amplitude for pulsations with $s \geqslant$ 99\% detected by {\itshape Chandra\/}.}
\label{hischa}
\end{figure}

\begin{figure}
\includegraphics[scale=0.45]{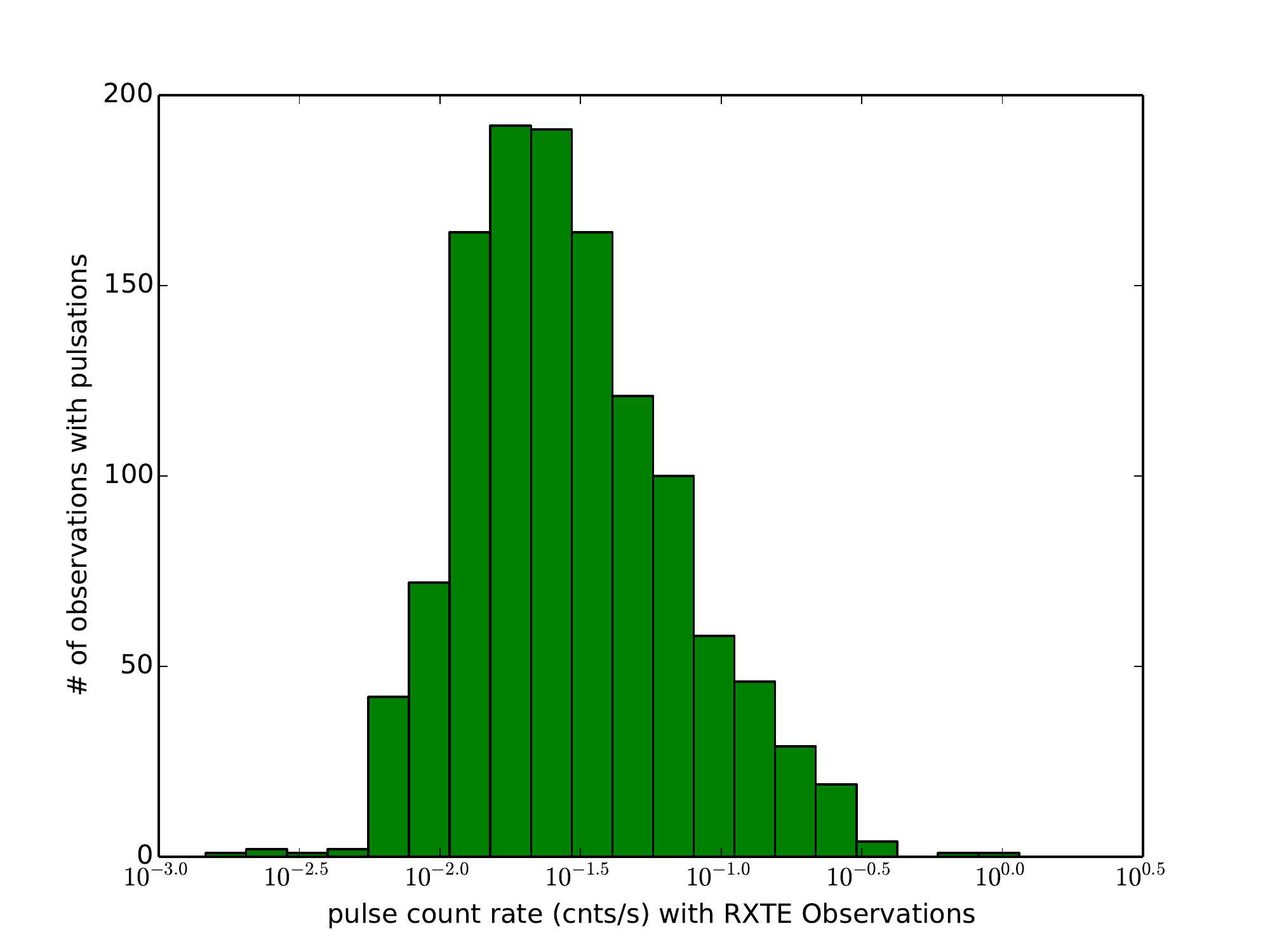}
\caption{Similar to Fig.~\ref{hisxmm}, showing the frequency histogram of pulse amplitude for pulsations with $s \geqslant$ 99\% detected by {\itshape RXTE\/}.   } 
\label{hisrxt}
\end{figure}

Figures~\ref{hisxmm}-\ref{hisrxt} show similar distributions for pulse amplitude 
for all three satellites (expressed in net count rate). They are the histograms of all known pulsars detected with $s\geqslant$ 99\% by \textit{XMM-Newton}, \textit{Chandra}, and \textit{RXTE}, respectively. The abscissae are equally binned in logarithmic count rate. It is apparent that a turnover occurs just above $\log_{10} A\simeq -1.2$ or 0.06 count/s for both ACIS and PN, and for the \textit{RXTE} amplitude histogram a similar turnover occurs at $\log_{10}A \simeq -1.7$ or 0.02 count/s. To compute the completeness fraction of the \textit{XMM-Newton}, \textit{Chandra} and \textit{RXTE} surveys the underlying pulsar $L_X$ and $PF$ distributions would need to be obtained first. We hope that modeling efforts will be motivated by statistical results such as these.  


Useful results can be inferred by transforming the empirical turnover points of Figs.~\ref{hisxmm}-\ref{hisrxt} into completeness limits in terms of flux or pulsar luminosity: Assuming the same PIMMS energy ranges and power-law model described in Section 2.4, the turnover points measured in Figs.~\ref{hisxmm}-\ref{hisrxt} yield completeness limits of flux 
$C_{XM}= 5.2 \times 10^{-13}$\flux, $C_{Ch}= 1.0 \times 10^{-12}$ \flux, $C_{RX}=2.5\times10^{-12} $ \flux, corresponding to luminosity values of $7.6\times 10^{35}$~erg/s, $1.6\times 10^{36}$~erg/s and $7\times 10^{36}$~erg/s, for \textit{XMM-Newton}, \textit{Chandra} and \textit{RXTE}, respectively.

\subsection {Time Evolution of Source Properties}
\label{sect:history}
In the combined pipeline, the standard data products for each pulsar include the pulse period,  luminosity, pulsed fraction, amplitude, and the spin period detection significance. The time evolution of these quantities yields the period derivative $\dot{P}$, the accretion torque, the orbital period, and the duty cycle of each source. Time-series plots of these quantities are provided as multi-panel PDF figures in the online journal for all pulsars in the sample, and the underlying data are provided in machine readable form. These time series can be read as a history of the on/off status and outburst state of each pulsar as a function of time. Such information is useful for comparison with other time-domain databases extending over the same time period, for example the OGLE and MACHO optical monitoring facilities. 
The plots are also intended as a useful resource for researchers interested in selecting their own sub-samples for further analysis or modeling.

As an example, the results for the HMXB pulsars SXP348 and SXP1323 are illustrated in Figs.~\ref{sxp348} and~\ref{sxp1323}, respectively. Each plot has 5 panels, with the data from each satellite plotted with a different color and symbol.  

The first (top) panel is the time series of $L_X$ in units of \lx~following the cross calibration scheme described in section~\ref{sect:calibration}. These are in fact the same points that appear in Fig.~\ref{fig:lumi}. $L_X$ values are plotted for all positive point-source detections (whether pulsed or not) from \chandra ~and \xmm, and for \rxte  ~pulsed detections.

The second panel reports the pulsed fraction for  \chandra ~and \xmm ~detections with pulsations at $s\geqslant99\%$ with solid symbols. Detections at $s < 99 \%$ are open symbols.

The third panel shows pulse amplitude in units of count rate, where the \chandra~and \xmm~rates are in units of PN count $\text{s}^{-1}$ after scaling as described in Section~\ref{sect:calibration}, and \rxte~rate is in units of count $\text{PCU}^{-1}\text{s}^{-1}$. We have followed the practice of earlier works (e.g.~\citealt{gal08}) by plotting \rxte~values even in cases of no pulsation detection; these open points are intended to represent upper limits. 

The fourth panel shows the spin period ($P$), with values plotted only for cases of detection at $s\geqslant99\%$. A linear fit is also displayed in order to highlight the long-term trend in period derivative. In the example of SXP348 (Fig.~\ref{sxp348}) two epochs of spin-up are seen, each lasting about 500 days, between which the pulsar returns to its long-term average value. In our second example, SXP1323 (Fig.~\ref{sxp1323}) there is an overall  long-term spin-up trend, with other apparently organized variations superimposed. This analysis was performed for all the pulsars with sufficient data to do so, and the slopes of these linear fits, their uncertainty, and the standard deviation of the points around the fit are collected together in Table~\ref{tab:pdot}. Here standard deviation implies how much the period of the pulsar varies on long timescales: a low standard deviation indicates that the data points tend to be close to the best fitting line of the set, while a high standard deviation indicates that the data points are spread out over a wider range of values around the best fitting line.

The fifth (bottom) panel reports the significance of the highest power-spectrum peak in the search-range for that pulsar (See section~\ref{sect:pulsesearch}). Pulsation searches were only performed for light curves with greater than 50 counts in the case of \xmm~ and \chandra. The solid symbols denote $s\geqslant99\%$ and in turn dictate how the points in other panels are plotted.

\subsection {Long-term Period Variations}
\label{sect:Pdot}
\begin{figure*}
\epsscale{1.00}
\plotone{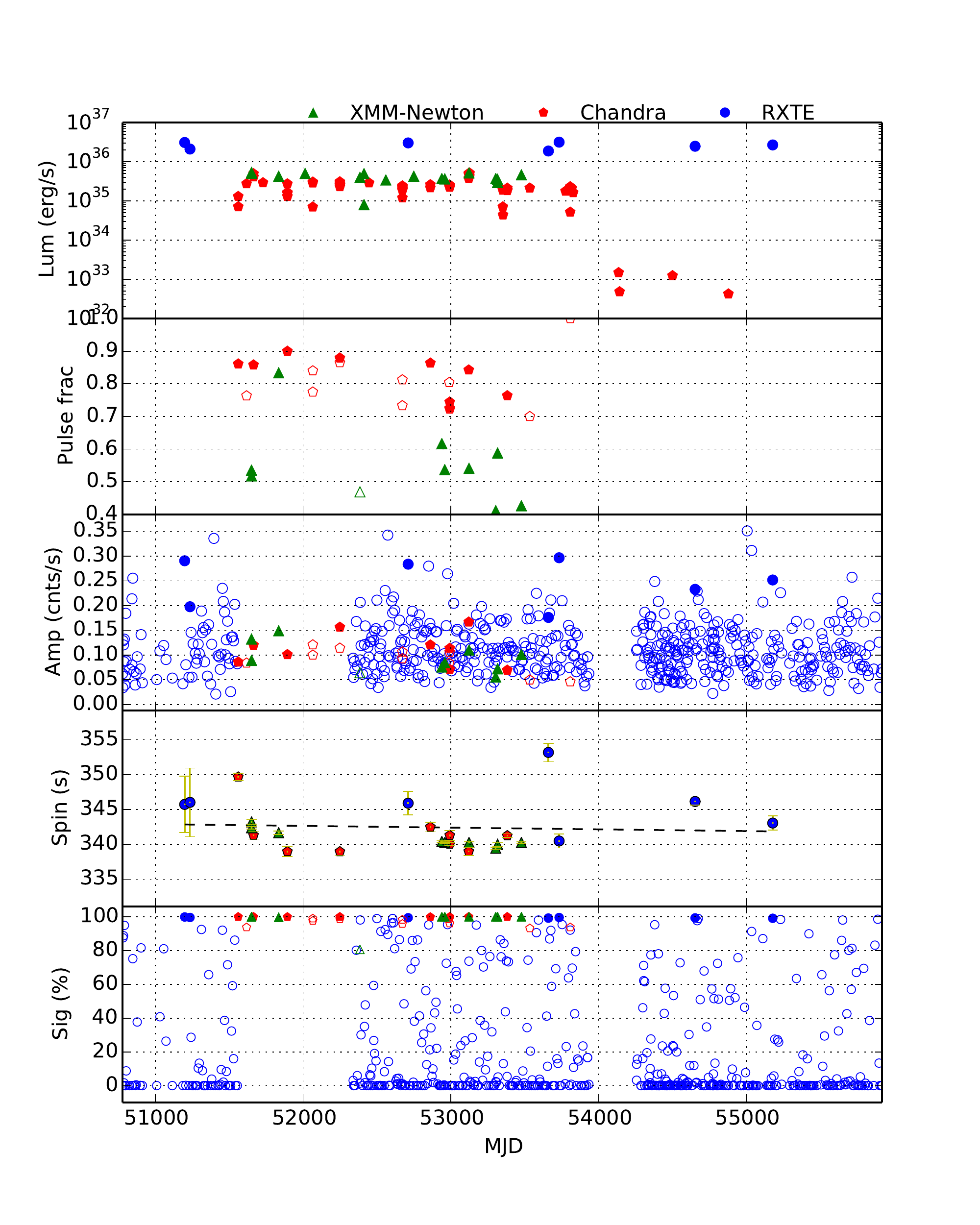}
\caption{Example multi-panel time-series plot for SXP348. The 5 panels (from top down) are luminosity ($L_X$), pulsed fraction, pulse amplitude (in units of count rate), spin period ($P_S$) and the related pulse detection significance ($s$). Plot symbols are green for {\itshape XMM-Newton\/}, red for {\itshape Chandra\/}, and blue for {\itshape RXTE\/}.  Solid symbols denote detections at $s\geqslant99\%$ for pulsed quantities. Open symbols denote $s<99\%$ for \xmm~ and \chandra. Open symbols for {\itshape RXTE\/} denote amplitude of variability in the lightcurve at the pulsar's period, it can be treated as the upper limit for pulse amplitude. The $L_X$ values for {\itshape XMM-Newton\/} and {\itshape Chandra\/} come from positive point-source detections, while \rxte~$L_X$ are derived from pulse amplitude. Full details are provide in the text (Section~\ref{sect:calibration}). Note, the fit value of the slope in the fourth panel are provided in Table~\ref{tab:pdot}.}
\label{sxp348}
\end{figure*}

\begin{figure*}
\epsscale{1.00}
\plotone{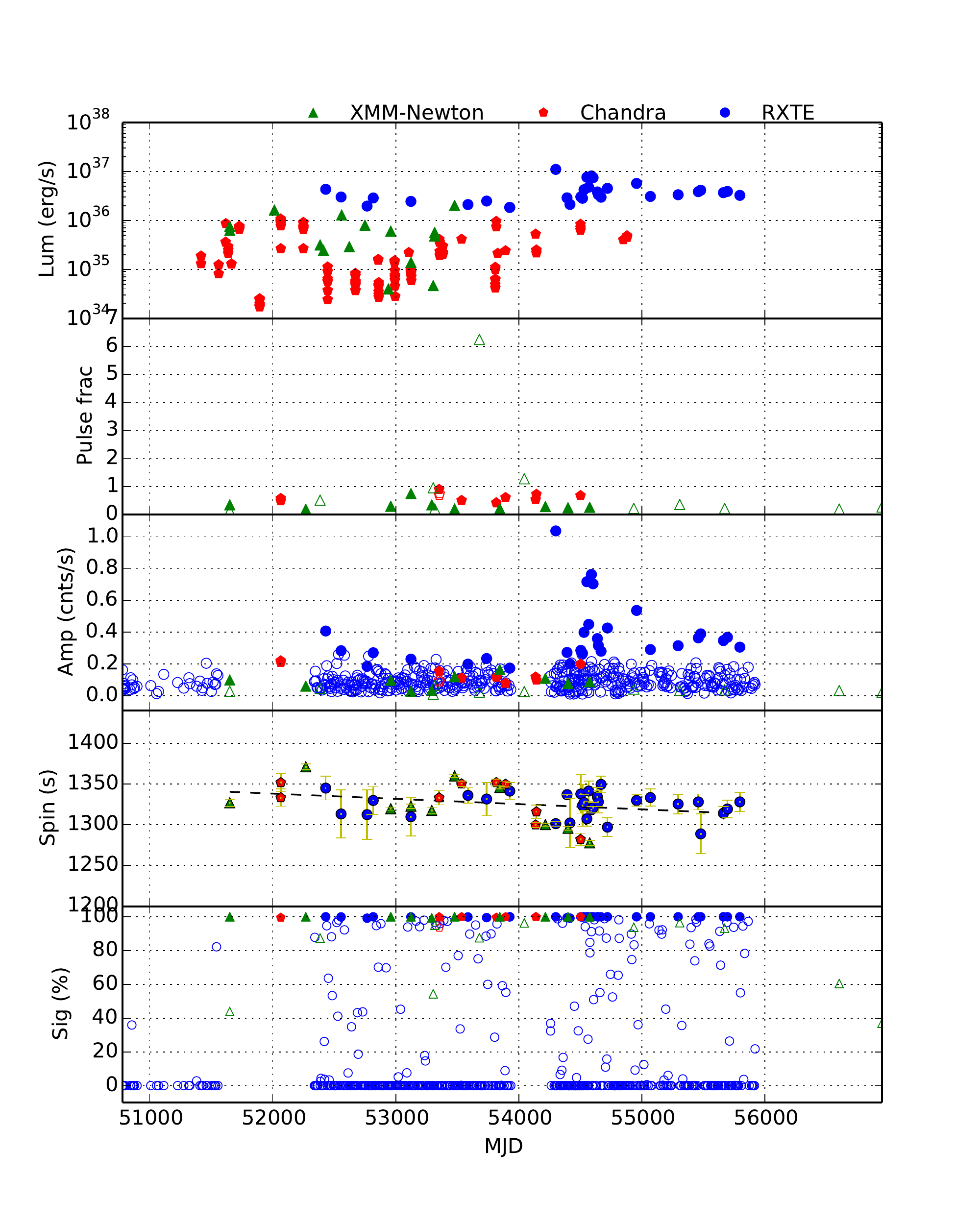}
\caption{Example multi-panel time-series plot for SXP1323. See the caption of Fig.~\ref{sxp348} for explanation.}
\label{sxp1323}
\end{figure*}

As expected for accreting pulsars in binary systems, all the  pulsars in our library show variations in their pulse periods.
These variations stem from periodic doppler shifting of the
frequency due to orbital motion and from the exerted accretion
torques. The first type of periodic variation may reveal the orbital period of the binary (see for example~\citealt{tow13}), whereas
the second type of secular variation gives us a window into the
physics of angular momentum transfer and dynamical evolution in
binary systems \citep{dav73,gho79}. 
The two types of variations can be seen in some of the multi-panel figures included in the online journal as supplementary material. The orbital modulation in some sources is not very clear, e.g., SXP 348 in Fig.~\ref{sxp348}, due to the few spin period data points. \cite{sch06} found weak evidence for an orbital period of 93.9 days for SXP 348 in OGLE-\RN{2} data. However, \cite{sch13} reported from re-analysis of the same data (plus OGLE-\RN{3} and -\RN{4} data in which it was not found), that the period was an artifact. \cite{coe15} simply quoted its orbital period as 93.9 days. From MJD 51000 until MJD 52000 is about 10 orbital cycles (with an orbital period of 94 days). During MJD 51000--52000, it only shows spin down and up once. SXP1323 (Fig.~\ref{sxp1323}) has no secure orbital period known according to \cite{coe15}. Here we carry out a linear regression
of the spin periods of 53 pulsars for which we have enough data
in order to search for secular drifts on timescales much longer than the orbital periods. 

We have investigated the entire spin-period history of each pulsar in our library by calculating the best-fit slope to the measured spin periods and its corresponding errors, as described in Section~\ref{sect:history}. The best-fit slopes, i.e., the measured $\dot{P}$ values, are listed in Table~\ref{tab:pdot} and they are also included in the multi-panel history plots such as Figs.~\ref{sxp348} and~\ref{sxp1323}. At the $\epsilon=1.5$ level or better, we find that 12 pulsars spin up and 11 pulsars spin down, while 30 pulsars appear to have $\dot{P}\approx 0$. The "C" in Table~\ref{tab:pdot} does not mean that the 30 pulsars do not go through spin-up and/or spin-down episodes, but that no net changes of significance are observed over the $\sim$15 year duration of the survey.  Some sources could well have undergone spin period changes that have averaged out over the survey period. In our dataset, 5 pulsars only have one single detection with $s\geqslant99\%$ pulsations found as shown in Table~\ref{tab:pdot}. The remaining 9 pulsars do not have significant pulsations detected at all.

\begin{figure}
\includegraphics[scale=.450]{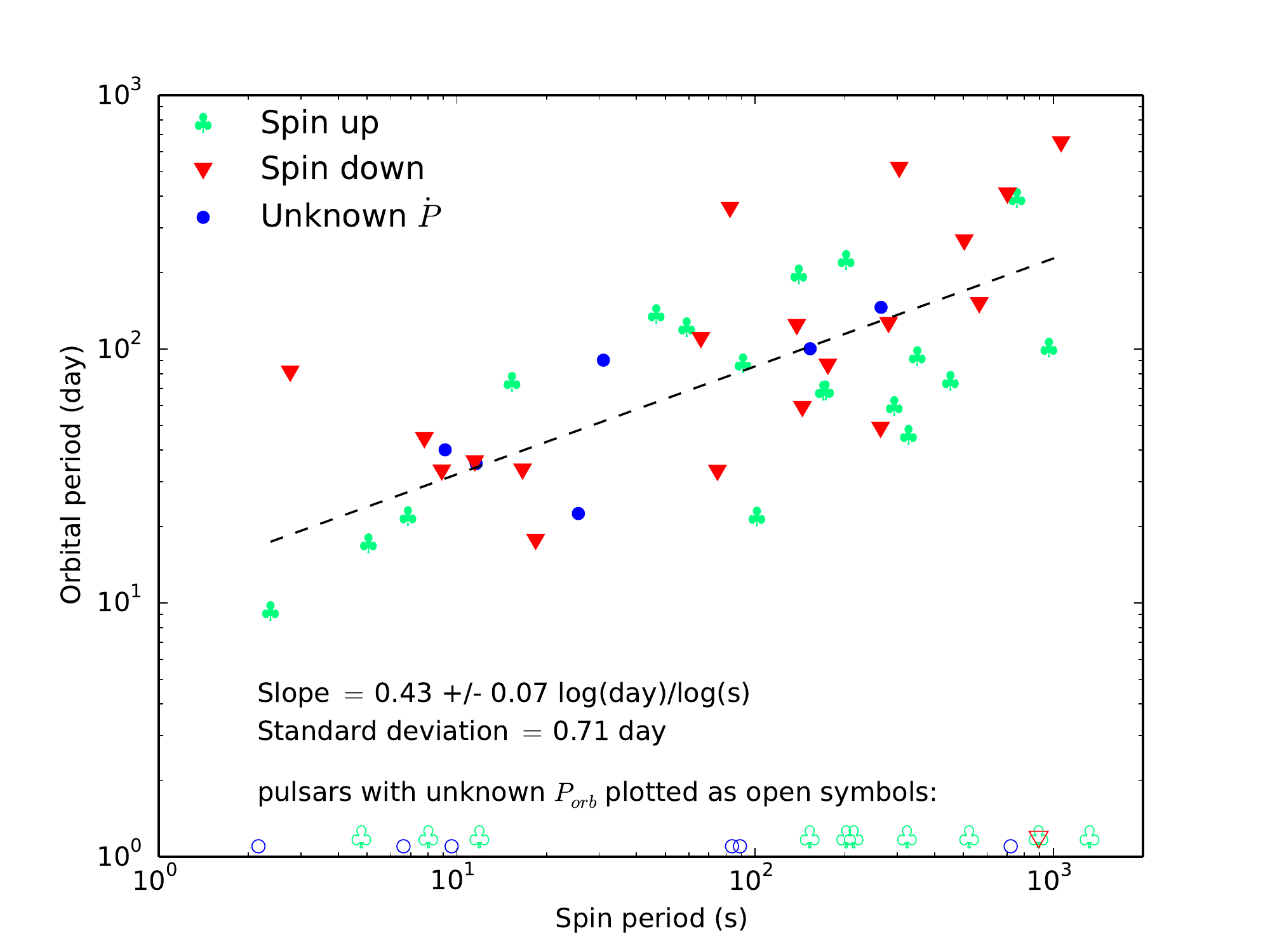}
\caption{Corbet diagram for all known pulsars in the SMC, with $\dot{P}$ information overlaid. Open and closed green (red) symbols show the pulsars that are spinning up (down) according to the values presented in Table~\ref{tab:pdot}. Blue symbols represent pulsars whose spin period derivatives are unknown due to lack of observations or lack of pulsations detected.}
\label{orbit}
\end{figure}

The cumulative outcome of this analysis is illustrated in Fig.~\ref{orbit} in which we plot the known orbital periods $P_{orb}$ versus the known spin periods $P$ \citep{coe15} with additional color-coded
information about the sign of $\dot{P}$. Green and red symbols
indicate that the pulsars spin up ($\dot{P}<0$) or down ($\dot{P}>0$), respectively (e.g., SXP1323, Fig.~\ref{sxp1323}). 
On the other hand, blue symbols indicate pulsars have no $\dot{P}$ value due to lack of observations or lack of pulsations detected in our survery. 
Unfilled symbols at the bottom of Fig.~\ref{orbit} show the pulsars for which $P_{orb}$ is unknown; currently, 43 SMC pulsars in our library do have measured orbital periods. 

The relation between the orbital and spin periods of 43 pulsars in Fig.~\ref{orbit} can be reasonably matched by a power law. The best-fit power law to the data is described by the equation 
\begin{equation}
P_{orb} = 12.066 \left(\frac{P}{1\ {\rm s}}\right)^{0.425}~~{\rm days}\ .
\label{cd}
\end{equation}
This equation shows that the longer the orbital period the wider the binary orbit, hence lower mass transfer with lower specific angular momentum occurs, and ultimately the pulsar rotation ends up being slower. Fig.~\ref{orbit} is effectively our updated version of the well-known \cite{corbet1984} diagram for the SMC. 
Overall, the HMXB spin period distribution has been shown to be bimodal by \cite{kni11}. Those authors suggested that there are two underlying populations tesulting from the two distinct types supernovae---electron capture and iron-core-collpase---that produce neutron stars.

\begin{deluxetable*}{l|lll|lll|ll}
\tablecaption{Summary of Individual Pulsar Observations and Detections}
\tablehead{ \colhead{Pulsar} &
 \colhead{XMM-Newton PN}           & \colhead{}  &  \colhead{} &   
\colhead{Chandra ACIS }  & \colhead{} & \colhead{}  & 
\colhead{RXTE PCA}   & \colhead{} \\
  \colhead{}                       & 
 \colhead{$N_{obs}$}  & \colhead{$N_{detect}$}  & \colhead{$N_{pulse}$} &  
 \colhead{$N_{obs}$}  & \colhead{$N_{detect}$}  & \colhead{$N_{pulse}$}   &
  \colhead{$N_{obs}$}  & \colhead{$N_{pulse}$} \\}
  
 \startdata                

SXP0.72   & 2 & 2 & 1 & 0 & 0 & 0 & 0 & 0 \\ 
SXP2.16 & 0 & 0 & 0 & 0 & 0 & 0 & 3 & 1   \\
SXP2.37 & 1 & 1 & 0 & 0 & 0 & 0 & 392 & 19 \\ 
SXP2.76   & 1 & 0 & 0 & 0 & 0 & 0 & 496 & 2 \\
SXP4.78 &0  & 0 & 0 & 0 & 0 & 0 & 804 & 11  \\
SXP5.05   & 2 & 0 & 0 & 5 & 1 & 0 & 889 & 5 \\
SXP6.85   & 1 & 1 & 0 & 0 & 0 & 0 & 616 & 52  \\
SXP6.88   & 4 & 3 & 0 & 4 & 1 & 0 & 0 & 0  \\ 
SXP7.78   & 1 & 1 & 0 & 3 & 1 & 0 & 804 & 28  \\
SXP7.92   & 8 & 0 & 0 & 4 & 2 & 1 & 879 & 5  \\
SXP8.02   & 7 & 7 & 1 & 15 & 8 & 1 & 559 & 2  \\ 
SXP8.8 & 0 & 0 & 0 & 0 & 0 & 0 & 861 & 43  \\ 
SXP8.88   & 1 & 1 & 0 & 3 & 2 & 0 & 0 & 0  \\
SXP9.13   & 4 & 4 & 0 & 7 & 7 & 0 & 650 & 1  \\ 
SXP11.5   & 1 & 1 & 0 & 0 & 0 & 0 & 599 & 10  \\ 
SXP11.6   & 1 & 1 & 0 & 0 & 0 & 0 & 0 & 0  \\ 
SXP11.9   & 5 & 1 & 1 & 1 & 0 & 0 & 394 & 2  \\
SXP15.3   & 4 & 4 & 0 & 3 & 2 & 0 & 655 & 16  \\ 
SXP16.6 & 0 & 0 & 0 & 0 & 0 & 0 & 650 & 14  \\ 
SXP18.3   & 5 & 5 & 1 & 4 & 2 & 1 & 854 & 73  \\
SXP22.1   & 3 & 1 & 0 & 0 & 0 & 0 & 3 & 0  \\
SXP25.5   & 3 & 3 & 0& 4 & 2 & 0 & 0 & 0  \\ 
SXP31.0   & 1 & 1 & 0 & 0 & 0 & 0 & 0 & 0  \\ 
SXP46.6   & 4 & 1 & 10& 4 & 1 & 10& 902 & 79  \\
SXP51 & 0 & 0 & 0 & 0 & 0 & 0 & 653 & 38  \\ 
SXP59.0   & 6 & 3 & 2 & 4 & 3 & 2 & 902 & 93  \\
SXP65.8   & 1 & 1 & 1 & 2 & 2 & 1 & 307 & 3 \\ 
SXP74.7   & 5 & 5 & 0& 4 & 2 & 0 & 854 & 21  \\ 
SXP82.4   & 2 & 2 & 0 & 4 & 1 & 0 & 863 & 19  \\ 
SXP91.1   & 2 & 1 & 0 & 0 & 0 & 0 & 784 & 60  \\ 
SXP95.2 & 0 & 0 & 0 & 0 &0  & 0 & 867 & 8  \\
SXP101 & 0 & 0 & 0 & 0 & 0 & 0 & 411 & 6 \\ 
SXP138   & 2 & 2 & 2 & 2 & 2 & 0 & 898 & 4 \\ 
SXP140   & 4 & 2 & 0 & 4 & 1 & 0& 879 & 7 \\
SXP144 & 0 & 0 & 0 & 0 & 0 & 0 & 561 & 51 \\ 
SXP152   & 8 & 6 & 1 & 2 & 2 & 1 & 561 & 19  \\
SXP153   & 1 & 1 & 1 & 23 & 2 & 0 & 296 & 0  \\ 
SXP169   & 2 & 2 & 2 & 0 & 0 & 0 & 505 & 33  \\ 
SXP172   & 3 & 3 & 2 & 3 & 3 & 0 & 863 & 45  \\ 
SXP175   & 6 & 6 & 0& 1 & 1 & 0 & 560 & 8  \\
SXP202A   & 13 & 10 & 0 & 0 & 0 & 0 & 567 & 14  \\ 
SXP202B   & 3 & 0 &0  & 0 &0  & 0 &  &   \\ 
SXP214   & 4 & 1 & 1 & 3 & 2 & 1& 854 & 16  \\ 
SXP264   & 4 & 4 & 1& 3 & 3 & 0 & 650 & 7  \\
SXP280   & 6 & 0 & 0 & 1 & 1 & 0& 555 & 6  \\ 
SXP292   & 3 & 3 & 0 & 4 & 1 & 0 & 644 & 10  \\
SXP293   & 6 & 3 & 0 & 0 & 0 & 0 & 944 & 13  \\ 
SXP304   & 20 & 19 & 4 & 31 & 4 & 0 & 557 & 7  \\ 
SXP323   & 3 & 3 & 2 & 4 & 4 & 2 & 649 & 22  \\ 
SXP327   & 1 & 1 & 0 & 3 & 3 & 2 & 827 & 5  \\
SXP342   & 3 & 3 & 1& 4 & 0 & 0 & 899 & 23  \\ 
SXP348   & 28 & 22 & 9 & 60 & 53 & 9 & 554 & 7  \\ 
SXP455   & 13 & 11 & 8 & 21 & 10 & 6 & 555 & 6  \\ 
SXP504   & 2 & 2 & 1 & 3 & 3 & 3 & 909 & 32  \\ 
SXP523   & 8 & 8 & 0 & 77 & 1 & 0 & 552 & 9  \\ 
SXP565   & 12 & 8 & 1 & 4 & 4 & 1 & 872 & 12  \\
SXP645   & 6 & 3 & 2 & 3 & 2 & 0 & 936 & 14  \\
SXP701   & 1 & 1 & 0 & 2 & 2 & 1 & 936 & 28  \\ 
SXP707   & 3 & 1 & 0 & 0 & 0& 0 & 0 & 0  \\ 
SXP726   & 29 & 22 & 2 & 52 & 3 & 2 & 299 & 5  \\ 
SXP756   & 3 & 3 & 2 & 4 & 3 & 2 & 391 & 23  \\
SXP893   & 3 & 3 & 1 & 7 & 7 & 5& 642 & 36  \\
SXP967   & 5 & 4 & 2 & 3 & 1 & 0 & 242 & 9  \\ 
SXP1062   & 6 & 6 & 6 & 11 & 11 & 9 & 0 & 0  \\ 
SXP1323   & 36 & 36 & 10 & 108 & 63 & 9 & 508 & 31  \\
SXP4693   &4   & 3 & 1 & 3 & 1 & 0 & 0 & 0  \\ 
\tableline 
TOTAL & 326 & 254 & 78 & 517 & 232 & 69 & 36316 & 1113  \\
\enddata
\tablecomments{Summary of Pulsar Library Observations and Detections tabulated by Satellite/Instrument and Pulsar.
For each instrument we report $N_{obs}$ = the number of ObsIDs containing the pulsar in the instrument FOV, $N_{detect}$= number of detections of the pulsar as a point source (imaging instruments only), and $N_{pulse}$= number of detections of the pulsar's spin period at $\geqslant$99\% significance.}
 \label{tab:pulsars} 
\end{deluxetable*}

\begin{table*} 
\centering
\caption{Long-term Average Spin Period Trends ($\dot{P}$)} 

\begin{tabular}{|c c c c c c|} 
\hline
Pulsar & $\dot{P}$    &   $\sigma_s$    &  $\epsilon$  & Class & $\#$ of\\ 
            &   (sec/day) &   (sec) &                     &             & points       \\
  \hline
SXP2.37 & -1.2e-05 $\pm$ 8e-06 & 0.010 & -1.50 & U&19\\
SXP2.76 & 3.0e-05 $\pm$ 0.0e-06 & 0.000 & $\gg1.5$ & D&2\\
SXP4.78 & -2e-06 $\pm$ 1e-06 & 0.004 & -2.00 & U&11\\
SXP5.05 & -6e-06 $\pm$ 1.2e-05 & 0.021 & -0.50 & C&5\\
SXP6.85 & -1e-06 $\pm$ 2e-06 & 0.011 & -0.50 & C&52\\
SXP7.78 & 6e-06 $\pm$ 1e-06 &-0.50  &  6.00& D&28\\
SXP7.92 & 1.9e-04 $\pm$ 1.1e-04 & 0.171 & 1.72 & D&6\\
SXP8.02 & -1.30e-05 $\pm$ 8.0e-06 & 0.024 & -1.63 & U&4\\
SXP8.80 & 3e-06 $\pm$ 1e-06 & 0.011 & 3.00 & D&43\\
SXP11.5 & 1.4e-05 $\pm$ 1.6e-05 & 0.040 & 0.875 & C&10\\
SXP11.9 & -1.2e-05 $\pm$ 4e-06 & 0.025 & -3.00 & U&3\\
SXP15.3 & -2e-06 $\pm$ 5e-06 & 0.015 & -0.40 & C&16\\
SXP16.6 & 2e-06 $\pm$ 3e-06 & 0.015 & 0.67 & C&14\\
SXP18.3 & 3e-06 $\pm$ 4e-06 & 0.025 & 0.75 & C&75\\
SXP46.6 & -4.8e-05 $\pm$ 1.6e-05 & 0.229 & -3.00 & U&79\\
SXP51.0 & 5.3e-05 $\pm$ 1.7e-05 & 0.155 & 3.12 & D&38\\
SXP59.0 & -4.4e-05 $\pm$ 1e-05 & 0.110 & -4.40 & U&93\\
SXP65.8 & 1.0e-04 $\pm$ 8.4e-05 & 0.029 & 1.19 & C&5\\
SXP74.7 & 7.7e-05 $\pm$ 1.3e-04 & 0.966 & 0.57 & C&21\\
SXP82.4 & 5.3e-05 $\pm$ 8.2e-05 & 0.283 & 0.65 & C&19\\
SXP91.1 & -1.144e-03 $\pm$ 7.3e-05 & 1.140 & -15.67 & U&60\\
SXP95.2 & 2.2e-04 $\pm$ 1.4e-04 & 0.234 & 1.56 & D&8\\
SXP101 & -1.6e-04 $\pm$ 3.4e-04 & 0.860 & -0.48 & C&6\\
SXP138 & 4.6e-04 $\pm$ 2.6e-04 & 0.577 & 1.75 & D&6\\
SXP140 & -1.1e-04 $\pm$ 5.0e-04 & 0.837 & -0.21 & C&7\\
SXP144 & 7.4e-04 $\pm$ 1.2e-04 & 0.950 & 6.24 & D&51\\
SXP152 & -1.06e-04 $\pm$ 7.2e-05 & 0.338 & -1.47 & C&21\\
SXP169 & -6.3e-04 $\pm$ 1.4e-04 & 1.010 & -4.52 & U&35\\
SXP172 & -3.03e-04 $\pm$ 7.0e-05 & 0.642 & -4.33 & U&47\\
SXP175 & 2.5e-04 $\pm$ 2.5e-04 & 0.639 & 1.01 & C&8\\
SXP202A & -2.9e-04 $\pm$ 2.7e-04 & 1.236 & -1.06 & C&14\\
SXP214 & -8.9e-05 $\pm$ 2.0e-04 & 1.135 & -0.45 & C&18\\
SXP264 & 1.4e-04 $\pm$ 6.9e-04 & 1.564 & 0.20 & C&8\\
SXP280 & 1.0e-03 $\pm$ 7.9e-04 & 1.168 & 1.27  & C&6\\
SXP292 & 2.7e-04 $\pm$ 3.4e-04 & 0.868 & 0.81 & C&10\\
SXP293 & -5.0e-05 $\pm$ 1.7e-04 & 0.774 & -0.29 & C&13\\
SXP304 & 2.9e-04 $\pm$ 8.7e-04 & 1.621 & 0.33 &C&11 \\
SXP323 & -1.51e-03 $\pm$ 2.8e-04 & 1.829 & -5.44 & U&27\\
SXP327 & -3.0e-04 $\pm$ 2.6e-04 & 0.449 & -1.18 & C&7\\
SXP342 & 1.48e-03 $\pm$ 9.1e-04 & 5.294 & 1.63 & D&24\\
SXP348 & -2.5e-04 $\pm$ 7.6e-04 & 3.531 & -0.33 & C&25\\
SXP455 & -1.3e-03 $\pm$ 9.5e-04 & 4.244 & -1.35 & C&20\\
SXP504 & 9e-06 $\pm$ 3.8e-04 & 3.713 & 0.02 &C &37\\
SXP523 & -1.8e-03 $\pm$ 2.1e-03 & 5.313 & -0.87 & C&9\\
SXP565 & 2.3e-04 $\pm$ 1.1e-03 & 4.497 & 0.20 & C&14\\
SXP645 & 1.5e-04 $\pm$ 6.4e-04 & 3.220 & 0.24 & C&16\\
SXP701 & 3.0e-03 $\pm$ 1.0e-03 & 7.288 & 2.93 & D&31\\
SXP726 & 2.0e-04 $\pm$ 6.2e-03 & 10.022 & 0.03 & C&9\\
SXP756 & -4.1e-03 $\pm$ 1.6e-03 & 10.226 & -2.48 & U&27\\
SXP893 & -6.8e-04 $\pm$ 1.2e-03 & 9.591 & -0.59 & C&42\\
SXP967 & -6.3e-04 $\pm$ 1.50e-03 & 5.452 & -0.43 & C&11\\
SXP1062 & 7.1e-03 $\pm$ 2.2e-03 & 2.889 & 3.23 & D&15\\
SXP1323 & -6.5e-03 $\pm$ 2.8e-03 & 19.014 & -2.32 & U&50\\
SXP0.72 & - & -  & - & - &1\\
SXP2.16 & - & -  & - & - &1\\
SXP9.13 & - & -  & - & - &1\\
SXP153 & - & - & - & -&1\\
SXP4693 & - & - & - & -&1\\
\bottomrule
\label{tab:pdot}
\end{tabular}
\tablecomments{Period derivative $\dot{P}$ and its error are determined from a linear fit to the time series of $P_{spin}$ measurements. $\sigma_s$ is the standard deviation indicating how much the period of the pulsar varies on long timescales around the linear fit. 
$\epsilon$ is the ratio of $\dot{P}$ to its own error as described in Section~\ref{sect:Pdot}. Classification is based on the sign and value of $\epsilon$: U: $\epsilon \leqslant -1.5$ (long-term spin up), D: $\epsilon \geqslant +1.5$ (long-term spin down), C: $-1.5<\epsilon <+1.5$ (consistent with no long-term variation). In the last column, $\#$ of points is the number of detected pulsations used in the linear fit.}
\end{table*}   

\subsection {Public Access to the Library}
\label{sect:access}

We will incrementally release this library to the public, from which one can download all of the light curves, periodograms, point source event lists, spectra (within the soft, hard and broad energy bands) of each known pulsar in the SMC \& LMC. For this paper,  release 1.0 provides the long-term observable parameters (i.e., count rate, luminosity,  spin period, pulsation amplitude, pulsed fraction) of the known pulsars in the SMC with the combination of all 3 satellite detections. The library is provided in catalog form and as a series of multi-panel PDF figures (58 in total).  
\\

\section{Discussion and Conclusions}\label{summary}

Using the entire archives of \xmm, \chandra, and \rxte~ observations up to the end of 2014, we have produced a comprehensive library of SMC X-ray pulsar properties. A  total of 37779 observations of 67 known pulsars in the SMC are found by our analysis pipelines. Of these observations, $\sim$1300 pulse period measurements (pulsation detection significance $\geqslant$99\%) were made (Table~\ref{tab:observations}). Histograms illustrating the contributions of each satellite are shown in Figs.~\ref{fig:xmm}-\ref{rxt}, and detailed listings are provided in Table~\ref{tab:pulsars}; the overall numbers are helpful as a guide for the amounts of each type of data product that can be found in the library.  The distributions of count-rate and pulse amplitude 
suggest that at least 200 counts are generally needed for reliable pulse detection, as well as subsequent spectral energy analysis at moderate resolutions.

The library spans $\sim$15 years and includes all key observational parameters generated by individual pipelines for the 3 X-ray telescopes which, in turn, are combined to produce a multivariate time-series for each X-ray pulsar (as illustrated in Figs.~\ref{sxp348} and~\ref{sxp1323}). These time series are provided in PDF and/or machine readable table form for each of the 67 SMC pulsars. 

One of these physical parameters, luminosity, as the function of the spin periods can map out the boundary of the pulsars in the Small Magellanic Cloud as $10^{31.2}$~to~$10^{38}$~erg~s$^{-1}$. The fast pulsars are less frequently observed. They are rarely experiencing outburst. Once they are in the outburst state, they are more likely to be the giant outburst. The very few detected fast pulsars tend to have a higher luminosities with \textit{XMM-Newton} observations. 
They might go through the outburst state, e.g., SMC X-2 \citep{pal16}, SMC X-3 \citep{tow17, wen17}. 
From the histogram Figs. 1-3, by chance, the long spin period pulsars are more frequently observed. The imbalance may guide us to propose the observing proposals for the fast pulsars.

For this paper introducing the library, we have performed an analysis of long-term period derivatives as an example of the types of investigation that can be done with this large sample of pulsars. 
The data show that long-lived accretion torques are present in about half of the sample. \cite{Okazaki2013} have pointed out that the standard picture of spin-up occurring only during transient outbursts is difficult to reconcile with the fact that the viscous timescale in the accretion disks is often longer than the orbital period. Nonetheless, the transient spin-up paradigm grew out of observations of the few pulsars with long-term period monitoring such as EXO 2030+375 \citep{Wilson2008}. By fitting the long-term pulse period values with linear models, we have determined that, at a level of $\epsilon=1.5$ or better, 12 SMC pulsars are spinning up and 11 spinning down on the $\sim$15 year timescale of the survey. On the other hand, 30 pulsars are consistent with no significant long-term average spin period changes and the remaining 14 pulsars remain uncategorized due to paucity of high signal-to-noise spin period measurements. The largest long-term average spin-up and spin-down values are -6.5e-03 $\pm$ 2.8e-03 s/day (SXP1323) and 7.1e-03 $\pm$ 2.2e-03 s/day (SXP1062), respectively. For comparison, such a very long-lived spin-up has been previously observed in just a few systems \citep[e.g.,][]{Krivonos2015}.
The relationship between the long-term spin-up rate and the X-ray luminosity has been the subject of a recent study \citep{Sugizaki2015} using the {\it Fermi} $\gamma$-ray observatory and the MAXI X-ray monitor on the International Space Station (ISS). This study found that $\dot{P}$ and $L_X$ are closely correlated ($-\dot{P} \propto{L_X}^{6/7}$), as was also found by \cite{coe10} for a sample of SMC pulsars. With this larger set of $\dot{P}$ values it will be possible to explore variations in the relationship predicted by \cite{gho79}.  

The initial public release of our library that accompanies this paper includes the long-term observable properties of the known pulsars in the  SMC. Future releases will include mid- and high-level data products for every individual pulsar detection. These products will include high time-resolution (raw and folded) light curves, periodograms, single source event lists, and calibrated energy spectra. We are also working to add to the library the 14 Be/X-ray pulsars in the LMC. As new observations enter the archive, our library will be updated and it will serve as a permanent resource for the community.  Furthermore, we note other contemporary data-mining efforts that are under way to {\it discover} pulsars in the archival data \citep{Israel2016}.

The ultimate goal of this project is to unleash the power of statistics on a large, unbiassed, observational sample of Magellanic Be/HMXB pulsars. These pulsars constitute precisely such a sample due to the presence of a large population of HMXBs within a small volume of space, at a known distance, that are also embedded in a readily resolvable stellar population. 
The rich statistical results from our library will certainly improve our understanding of magnetized accretion processes in HMXBs as the various products (pulse profiles, periodograms, X-ray spectra) will be used in the testing of theoretical models of gas flows through NS magnetospheres.

\section*{Acknowledgments}
We thank J. Homan for his assistance with the \textit{XMM-Newton} data reduction, and W. C. G. Ho and J. S. Hong for their valuable comments which improved the quality of the paper. We also thank the anonymous referee whose helpful suggestions have improved the paper. This work was supported by NASA grant NNX14-AF77G.

\appendix
The 58 multi-panel figures are at \url{https://authortools.aas.org/AAS03548/FS9/figset.html}.

\end{document}